\documentclass[a4paper,fleqn,usenatbib]{mnras}
\usepackage[T1]{fontenc}
\usepackage{ae,aecompl}
\usepackage{natbib}
\usepackage{amssymb}
\usepackage{amsmath}
\usepackage{mathptmx}
\usepackage{epstopdf}
\usepackage{booktabs}
\usepackage{float}
\usepackage{subfig}
\usepackage{soul}
\usepackage{epsfig,color,ulem}
\usepackage{tabularx}
\usepackage{multirow}
\usepackage[dvipsnames]{xcolor}
\definecolor{new_color}{rgb}{0.0, 0.6, 0.7}

\newcommand{\Eiso}{E_{\rm{iso}}}
\newcommand{\A}{{\it LH}}
\newcommand{\B}{{\it LM-4}}
\newcommand{\C}{{\it LM-2}}
\newcommand{\Ce}{{\it LM-2e}}
\newcommand{\D}{{\it LM-1}}
\newcommand{\E}{{\it LHpn}}
\newcommand{\F}{{\it LM-2w}}
\newcommand{\G}{{\it SH}}
\newcommand{\h}{{\it SM-2}}

\newcommand{\msun}{\,{\rm M_{\odot}}}
\newcommand{\cm}{\,{\rm cm}}

\newcommand{\mx}{\,{\rm max}}
\newcommand{\erg}{\,{\rm erg}}

\newcommand{\s}{\,{\rm s}}
\newcommand{\LogT}{{\rm Log}_{_{10}}}
\def\gsim{ \lower .75ex \hbox{$\sim$} \llap{\raise .27ex \hbox{$>$}} }
\def\lsim{ \lower .75ex\hbox{$\sim$} \llap{\raise .27ex \hbox{$<$}} }
\def\app#1#2{%
	\mathrel{%
		\setbox0=\hbox{$#1\sim$}%
		\setbox2=\hbox{%
			\rlap{\hbox{$#1\propto$}}%
			\lower1.1\ht0\box0%
		}%
		\raise0.25\ht2\box2%
	}%
}

\title[The structure of magnetized GRB jets]{The structure of weakly-magnetized $ \gamma $-ray burst jets}

\author{
	Ore Gottlieb \altaffilmark{1}, Omer Bromberg \altaffilmark{1}, Chandra B. Singh \altaffilmark{1}, Ehud Nakar \altaffilmark{1}
}
\author[O. Gottlieb et al.]{
	Ore Gottlieb$^{1}$\thanks{oregottlieb@mail.tau.ac.il},
	Omer Bromberg$^{1}$,
	Chandra B. Singh$^{1,2}$,
	Ehud Nakar$^{1}$
	\\
	$^{1}${School of Physics and Astronomy, Tel Aviv University, Tel Aviv 69978, Israel}\\
	$^{2}${South-Western Institute for Astronomy Research, Yunnan University, University Town, Chenggong, Kunming 650500, People's Republic of China}
}

\pubyear{2020}
\begin{document}
	\label{firstpage}
	\pagerange{\pageref{firstpage}--\pageref{lastpage}}
	\maketitle	
	\begin{abstract}
		The interaction of gamma-ray burst (GRB) jets with the dense media into which they are launched promote the growth of local hydrodynamic instabilities along the jet boundary. In a companion paper we study the evolution of hydrodynamic (unmagnetized) jets, finding that mixing of jet-cocoon material gives rise to an interface layer, termed jet-cocoon interface (JCI), which contains a significant fraction of the system energy. We find that the angular structure of the jet + JCI, when they reach the homologous phase, can be approximated by a flat core (the jet) + a power-law function (the JCI) with indices that depend on the degree of mixing. In this paper we examine the effect of subdominant toroidal magnetic fields on the jet evolution and morphology. We find that weak fields can stabilize the jet against local instabilities. The suppression of the mixing diminishes the JCI and thus reshapes the jet's post-breakout structure. Nevertheless, the overall shape of the outflow can still be approximated by a flat core + a power-law function, although the JCI power-law decay is steeper. The effect of weak fields is more prominent in long GRB jets, where the mixing in hydrodynamic jets is stronger. In short GRB jets there is small mixing in both weakly magnetized and unmagnetized jets.  This result influences the expected jet emission which is governed by the jet's morphology. Therefore, prompt and afterglow observations in long GRBs may be used as probes for the magnetic nature at the base of the jets.
		
	\end{abstract}
	
	
	\section{Introduction}
	\label{sec:introduction}	
	
	Gamma-Ray Bursts (GRBs) are created by ultra relativistic jets launched from the vicinity of compact objects. Once launched, a GRB jet must drill through a dense medium surrounding the compact object: a stellar envelope \citep{MacFadyen1999} in the case of long GRBs (lGRBs) or, presumably, a neutron star (NS) merger ejecta \citep{Eichler1989,Narayan1992} in the case of short GRBs (sGRB). As the jet pushes through the medium, two shocks are formed: a bow shock that propagates into the medium and a reverse shock that forms at the jet's head. Ambient matter that crosses the bow shock, together with  jet material that crosses the reverse shock and spills sideways, form a hot cocoon that engulfs the jet and collimates it \citep[see e.g.][]{MacFadyen2001,Ramirez-Ruiz2002,Zhang2003,Morsony2007,Mizuta2009}.
	
	The jet collimation is obtained through an oblique shock that forms in the relativistic flow close to the jet base and gradually converges to the jet axis \citep{Bromberg2011b}. 
	Above the convergence point a series of repetitive weaker recollimation shocks form, which facilitate the collimated flow in the jet. The collimation reduces the cross section of the jet head, thereby accelerating its propagation through the dense medium. A fast head is essential for the successful breakout of the jet from the confining medium during the lifetime of the jet engine. Without strong collimation, the jet head would remain buried deep in the medium when the engine dies, and jets would fail to break out and produce luminous GRBs \citep{Lazzati2005a,Bromberg2011}.
	The interaction with the  cocoon may also slow down the fast head propagation by producing instabilities on the boundary separating the jet and the cocoon. If grown to large enough amplitudes they can lead to a substantial entrainment of baryonic matter into the jet, thereby reducing the specific enthalpy of the jet. This effect may considerably alter the jet dynamics and its emission properties (e.g. \citealt{Gottlieb2019b}).
	
	In Newtonian systems, whenever a light fluid accelerates onto a heavy one, the conditions across the contact discontinuity hold $\nabla\rho\nabla p<1$, where $\rho$ and $p$ are the mass density and pressure of the fluids on both sides of the shock. This state is unstable for the ``fingering" Rayleigh-Taylor instability (RTI, \citealt{Rayleigh1882,Taylor1950}), which induces mixing between the two fluids. In the case of collimated relativistic jets, \citet{Matsumoto2013a,Matsumoto2013} showed that RTI can grow on the jet-cocoon boundary surface, above the collimation point if the jet material is ``relativistically denser" than that of the cocoon. They introduced a relativistic stability criterion in which $\rho$ is replaced by $\rho h \Gamma^2$, the enthalpy density in the frame of the cocoon, where $h$ and $\Gamma$ are the specific enthalpy and Lorentz factor of the flow.  
	\citet{Matsumoto2017} carried out a linear stability analysis under the approximation of a planar interface moving perpendicular to the jet flow, and confirmed this criterion while adding a numerical factor of order unity to the equation. In their model the collimation of the jet drives a lateral acceleration towards the jet axis. If the jet material has a higher enthalpy density than the cocoon, the acceleration of cocoon material onto the jet generates RTI on the jet-cocoon boundary surface \citep{Matsumoto2019}.
	Once the collimation shock converges to the axis it reflects back towards the contact discontinuity on the jet boundary and induces a second instability, the impulsive Richtmyer-Meshkov instability (RMI, \citealt{Richtmyer1960,Meshkov1969}). RMI accelerates the growth of the RTI fingers and enhances the mixing between the two fluids. Ultimately the jet loses its coherent structure as the consecutive recollimation shocks are destroyed by the two types of instabilities. A movie demonstrating this process can be found \href{http://www.astro.tau.ac.il/~ore/instabilities.html}{here}\footnote{ \url{http://www.astro.tau.ac.il/~ore/instabilities.html}}.
	
	The mixing between the jet and the cocoon increases the baryon load in the jet. Consequently, the propagation velocity of the jet head drops and the breakout time grows. This behavior has been found in a variety of 3D relativistic hydrodynamic simulations of jets in a dense media \citep{Zhang2003,Rossi2008,LopezCamara2013,Harrison2018,Gottlieb2018a,Gottlieb2018b,Matsumoto2019}.
	In a parallel project \citep[][, hereafter GNB20]{Gottlieb2020b} we conduct a thorough study of the effects of the RTI and RMI on 
	hydrodynamic (unmagnetized) GRB jets propagating in different media. 
	We show that the instabilities, which grow on the contact discontinuity between the jet and the cocoon, generate a new layer of mixed material, which separates pure jet material from cocoon material. We term this layer as the {\it jet-cocoon interface} (JCI). 
	We study the properties of JCI and its effects on the observed emission in various jet configurations relevant for lGRBs and sGRB.
	We generally distinguish between two evolutionary stages, before and after the jet breaks out from the medium. 
	In the pre-breakout stage the instabilities grow along the jet boundary between the collimation point and the jet head, inciting mixing in the entire region. The degree of mixing in this stage and the amount of baryon loading  depend on the interplay between the jet and the cocoon and it can vary quite substantially with the system properties. 
	We find that the jet power, the injection angle and the medium density have the strongest effect on the degree of mixing.
	High power jets with small injection angles propagating in low density media develop fast moving heads. They are least affected by their cocoons and show a small degree of mixing. Wide angle, low power jets in high density media show stronger mixing and develop more prominent JCIs. 
	
	Once the jet breaks out of the medium, its head accelerates and disperses. The cocoon, which breaks out with the jet stops receiving fresh energy and gradually depressurizes. 
	The depletion rate is slow enough to keep the cocoon pressure inside the medium high for a long time with respect to the jet breakout time. During that time the jet-cocoon interaction remains strong and vigorous instabilities continue to grow along the jet boundary, albeit confined to grow only inside the medium.
	Once the jet material exits the medium, the mixing stops and the composition freezes out.
	
	The JCI that surrounds the jet core may considerably alter the emission of GRB jets (GNB20). So far the jet boundary instabilities and the properties of the JCI were only studied in the context of hydrodynamic jets. 
	The launching of the jets likely involves magnetic fields \citep{Blandford1977,Komissarov2001}, and presumably at least some of the field survives the journey along the jet \citep{Lyutikov2005,Metzger2010,Bromberg2016}. Since magnetic fields are known to suppress the growth of RTI \citep[e.g.][]{Chandrasekhar1961,Jun1995,Millas2017,Matsumoto2019}, their presence may alter the structure of the JCI and through that the observed emission as well. Understanding how the structure and properties of the jet-cocoon system change with the jet magnetization may help constraining the strength of the jet magnetic field from observations.
	
	In this work we focus on weakly-magnetized jets, where the magnetic fields are subdominant, i.e. we require that in the plasma frame $\sigma\equiv\frac{b^2}{4\pi \rho h}\ll1$, where $b$ is the plasma proper magnetic field and $\rho h$ is the proper enthalpy density. In the weak field regime magnetic fields can suppress the growth of RTI on the jet boundary while avoiding current-driven instabilities, such as the kink instability, which deform the entire jet body and alter its dynamics \citep{Baty2002,Nakamura2004,Giannios2006,Meliani2009,Mizuno2009,Mizuno2012,Bromberg2016,Tchekhovskoy2016,Kim2017}. We conduct our study using 3D relativistic magneto-hydrodynamic (RMHD) simulations that follow the jet prior and after its breakout from the medium. We monitor the structure and composition of the JCI and evaluate its effects on the emission seen by observers from different line of sights.

	The outline of the paper is as follows.
	In \S\ref{sec:models} we introduce the models considered here and the numerical setup of the simulations. In \S\ref{sec:mixing} we present the numerical results and compare the mixing of hydrodynamic (unmagnetized) jets and weakly-magnetized jets. In \S\ref{sec:post_breakout} we examine the evolution of weakly magnetized jets in the post-breakout phase, deduce the terminal structure of the jet-cocoon system and discuss its effect on the observed emission. We make a comparison between these results and the results from hydrodynamic jets. In \S\ref{sec:conclusions} we summarize and conclude.
	
	\section{Models and Setup}
	\label{sec:models}
	
	We study the jet evolution using 3D simulations with the PLUTO code \citep{Mignone2007}. The code has the advantage of having a flexible numerical scheme. Our integration setup includes a third order Runge-Kutta time stepping, piece-wise parabolic reconstruction with harmonic limiter, and an HLL Riemann solver. In order to avoid nonphysical states, slope-limited reconstruction with the MinMod limiter is adopted to handle shocks, and we use constrained transport to enforce $ \nabla \cdot {\bf B} =0$. We use an ideal equation of state with an adiabatic index 4/3, which is appropriate since the shocks are all radiation mediated and the shocked plasma energy density is dominated by radiation \citep[for a review of radiation mediated shocks see][]{Levinson2019}.
	We simulate both hydrodynamic and magnetized jets in lGRB as well as in sGRB configurations, and examine the effects of a toroidal magnetic fields on the evolution of the jets. 
	
	Our lGRB jet setup is based on the simulations run in \citet{Gottlieb2019b}. We use model $Lc$ from their work (termed here  $ \A $) for our canonical hydrodynamic jet. To study the stabilization effects of magnetic fields, we inject a toroidal field at the jet base and vary its strength between simulations.
	The setup includes a static, non-rotating stellar envelope with a mass of $ M = 10\msun $, a radius of $ R_\star = 10^{11}\cm $ and a density profile
	\begin{equation}\label{eq:star}
	\rho_a(r) = \rho_0 r^{-2}\Bigg(\frac{R_\star-r}{R_\star}\Bigg)^3~,
	\end{equation}
	where $\rho_0 = \frac{2}{\pi}\times 10^{23}~{\rm g~cm^{-1}}$.
	At time $ t = 0 $ we inject a hot jet with a specific enthalpy $h_{_0}\gg1$, a total (two sided) power of $ L_j = 10^{50}\erg~\s^{-1} $ and an initial Lorentz factor $ \Gamma_0 = 5 $. The jet can reach a maximal terminal four-velocity of $ u_{\infty,\mx} \equiv \Gamma_0h_0-1 = 500 $ if no mixing takes place.  
	The jet is injected continuously throughout the entire duration of the simulation as a cylinder with a radius $ r_{j,0} = 10^8\cm $ having a flow velocity aligned with the jet axis 
	(the $\hat{z}$ direction). The hot jet spreads quickly to a conical shape with an initial half opening angle $\theta_{j,0} \approx 0.7\Gamma_0^{-1} $ \citep{Mizuta2013,Harrison2018}. Consequently, we set the injection height (the location of the lower $\hat{z}$ boundary) at $ z_{0} = r_{j,0}/\theta_{j,0} $.
	To avoid a sharp jump across the jet boundary, which can result in numerical errors, the jet parameters are injected with a smooth profile
	$ \rm{cosh}^{-1}\big(\frac{r}{r_{j,0}}\big)^{\beta} $, where we use $ \beta = 8 $ for the hydrodynamic quantities of the jet (mass density, pressure and velocity), similar to the hydrodynamic jets in \citet{Gottlieb2019b}, and $ \beta = 6 $ for the magnetic fields.
	The injected magnetic field has a cylindrical radial profile that follows \citet{Mignone2009,Mignone2013}:
	\begin{equation}
	b_\phi = \sqrt{\frac{2\pi\rho_j\Gamma_0^2\sigma_0}{1-4\rm{log}\mathnormal{\frac{1}{2}}}}{\left\{\begin{array}{c} r/a \qquad\qquad\qquad\qquad\qquad r<a \\ \frac{a}{r}\Big(1-\frac{(r-a)^2}{(r_{j,0}-a)^2}\Big) \qquad a<r<r_{j,0} \end{array}\right\}}~,
	\end{equation}
	where $ a = r_{j,0}/2$ and $\sigma_0$ is the flow magnetization at $r=a$ and it is roughly the magnetization of the jet material prior to any mixing that may occur.
	For our studies we choose values of $ \sigma_0 = 10^{-1}, 10^{-2}, 10^{-4} $. See Table \ref{table} for the full list of the models used in this work.
	In addition to the canonical setup (used in models $\A, \B, \C, \D$), we investigate the effect of the opening angle on the stability of the jet.
	We carry out a single simulation of a magnetic jet with a wide opening angle (model $ \F $) and one simulation of a high power hydrodynamic jet with a narrow opening angle relative to the canonical opening angle of $8^\circ$ (model $ \E $).
	
	
	We examine the stabilization effects of magnetic fields in sGRB jets as well (models $\G$ and $\h$ in Table \ref{table}, which simulate a hydrodynamic jet and a magnetized jet with $\sigma_0=10^{-2}$, respectively).
	We consider a sGRB formed in the aftermath of a binary NS merger \citep[for review see][]{Nakar2019}. The merger is accompanied by an ejection of a few percent of solar mass from the system, which expands homologously \citep[][and references therein]{Nakar2019,Shibata2019}.
	As we learned from GW170817, a relativistic jet is launched following the merger, most likely after a short delay ($\lesssim 1$ s), possibly due to a delayed collapse of the merger product to a black hole  \citep[][and references therein]{Nakar2019}. The jet needs to break out of the ejecta before it can generate the GRB. We use the same ejecta model as in \citet{Gottlieb2019b}.
	The ejecta is composed of a core part having a mass $ M_c \approx 0.05 \msun $, expanding homologously with a non-relativistic velocity ($ v_c < 0.2c $) and a profile density $ \rho(r)=2.2\times10^{21}\rm{g~cm^{-1}}r^{-2} $. It is embedded in a ``tail" of light material moving at a mildly relativistic velocity with a steep density profile $ \rho(r) \propto r^{-14} $. The jet is injected with a delay of $0.6$s from the onset of the simulation, which marks the time of the merger. The canonical sGRB jet is injected with a luminosity: $ L_j = 2 \times 10^{49} {\rm ~ erg/s}$, an initial Lorentz factor $ \Gamma_0 = 5 $, and an asymptotic 4-velocity $ u_{\infty,\mx} = 500 $. The magnetized jet has the same hydrodynamic properties, and it carries a magnetic field with the same profile as in our magnetic lGRB models and $\sigma_0=10^{-2}$.
	See Table \ref{table} for the full list of parameters used in this case.
	
	\begin{table}
		\setlength{\tabcolsep}{2.4pt}
		\centering
		\begin{tabular}{ | l | c  c  c  c  c  c  c | }
			\hline
			lGRBs & $ \sigma_0 $ & $ L_j [10^{50}\rm{\frac{erg}{s}}] $ & $ \theta_{j,0} $ & $ u_{\infty,\mx} $ & $ t_b $ [s] & $ z_{\rm max} $ [cm] & $ r_{h,{\rm max}} $ [cm]\\ \hline
			$ \A $ & 0 & 1.0 & 0.14 & 500 & $ 15^\dagger $ & $ 10^{12} $ & $ 10^{12} $\\
			$ \B $ & $ 10^{-4} $ & 1.0 & 0.14 & 500 & 14 & $ 10^{11} $ & $ 10^{11} $\\
			$ \C $ & \multirow{2}{*}{$10^{-2}$} & \multirow{2}{*}{1.0} & \multirow{2}{*}{0.14} & \multirow{2}{*}{500} & \multirow{2}{*}{7} & $ 10^{11} $ & $ 4\times 10^{11} $\\
			$ \Ce $ & & & & & & $ 2\times 10^{11} $ & $ 2\times 10^{11} $\\
			$ \D $ & $ 10^{-1} $ & 1.0 & 0.14 & 500 & 5 & $ 10^{11} $ & $ 10^{11} $\\
			$ \E $ & 0 & 7.0 & 0.07 & 1000 & 6 & $ 8\times 10^{11} $ & $ 10^{12} $ \\
			$ \F $ & $ 10^{-2} $ & 1.0 & 0.24 & 300 & 15 & $ 10^{11} $ & $ 4\times 10^{11} $ \\
			\hline
			sGRBs & $ \sigma $ & $ L_j [10^{49}\rm{\frac{erg}{s}}] $ & $ \theta_{j,0} $ & $ u_{\infty,\mx} $ & $ t_d, t_b $ [s] & $ z_{\rm max} $ [cm] &  $ r_{h,{\rm max}} $ [cm]\\ \hline
			$ \G $ & 0 & 2.0 & 0.14 & 500 & 0.6; 1.4 & $ 8\times 10^{10} $ & $ 10^{11} $\\
			$ \h $ & $ 10^{-2} $ & 2.0 & 0.14 & 500 & 0.6; 1.4 & $ 8\times 10^{10} $ & $ 10^{11} $\\ \hline
		\end{tabular}
		\hfill\break
		
		\caption{The configurations of the simulations. $ \sigma_0=\frac{b^2}{4\pi h\rho} $ is the initial magnetic to thermal energy fluxes ratio, $ L_j $ is the total jet luminosity (two sided), $ \theta_{j,0} $ is the jet half-opening angle at the base, $ u_{\infty,\mx} = \Gamma_0h_0-1 $ is the terminal four velocity of the jet if it does not mix, defined by the initial Lorentz factor $ \Gamma_0 $ and the initial specific enthalpy $ h_0 $, $ t_d, t_b$ are the delay time and the breakout time, respectively, $ z_{\rm max} $ is the upper $ \hat{z} $ boundary of the grid, and $ r_{h,\rm max} $ is the jet head location at the end of the simulation. In cases where $r_{h,\rm max}>z_{\rm max}$ we allow the jet head to leave the box and continue the simulation until it reaches the location marked in the relevant column. In the sGRB models all times are measured from the time of the merger and $ t_b $ refers to the breakout from the core ejecta (see text for details).
			\newline
			$ ^\dagger$\scriptsize{The breakout time is shorter than the value in \citet{Gottlieb2019b} due to a higher resolution grid.}
		}
		\label{table}
	\end{table}
	
	Magnetic jets must be simulated with a higher resolution than hydrodynamic jets, to properly follow the MHD flow. The grid setup of the hydrodynamic jets is given in GNB20. The grids of the magnetic runs include three patches on the $ \hat{x} $ and $ \hat{y} $ axes independently and one on the $ \hat{z} $-axis.
	For the lGRB grid the inner patch on the $ \hat{x} $ and $ \hat{y} $ axes is uniform with 400 cells in the inner $ |2.5\times 10^9|\cm $. The outer patches are logarithmic with 80 cells in each direction from $ |2.5\times 10^9|\cm $ up to $ |10^{11}|\cm $. On the $ \hat{z} $-direction we use 1000 uniform cells from $ z_{0} $ to $ z_{\rm max}=10^{11} $ cm (in simulation $ \Ce $, which extends to $ z_{\rm max} = 2R_\star $ we use 2000 cells). In total we have $ 560\times 560\times 1000 (2000) $ cells.
	For the sGRB grid the inner patch on $ \hat{x} $ and $ \hat{y} $ axes is uniform with 160 cells in the inner $ |10^9|\cm $, and the outer patches are logarithmic with 400 cells on each direction up to $ |10^{11}|\cm $. On the $ \hat{z} $-axis there are 1500 uniform cells from $ z_{0} $ to $ 1.2\times 10^{11}\cm $. In total we have $ 960\times 960\times 1500 $ cells.
	We present convergence tests in Appendix \ref{sec:convergence}.
	
	\section{propagation \& stability before breakout}
	\label{sec:mixing}
	
	\begin{figure*}
		\centering
		\includegraphics[scale=0.39]{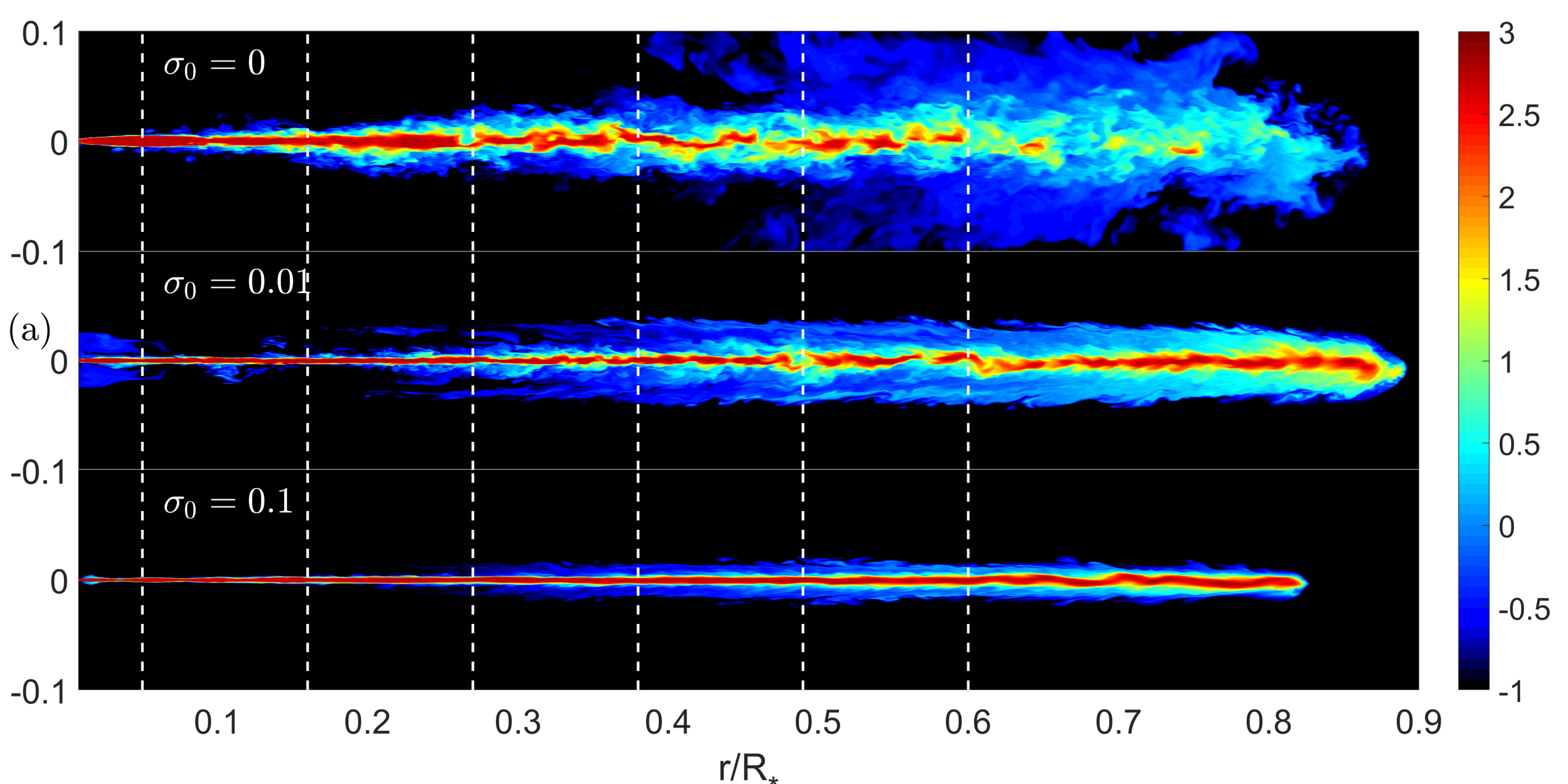}
		\includegraphics[scale=0.39]{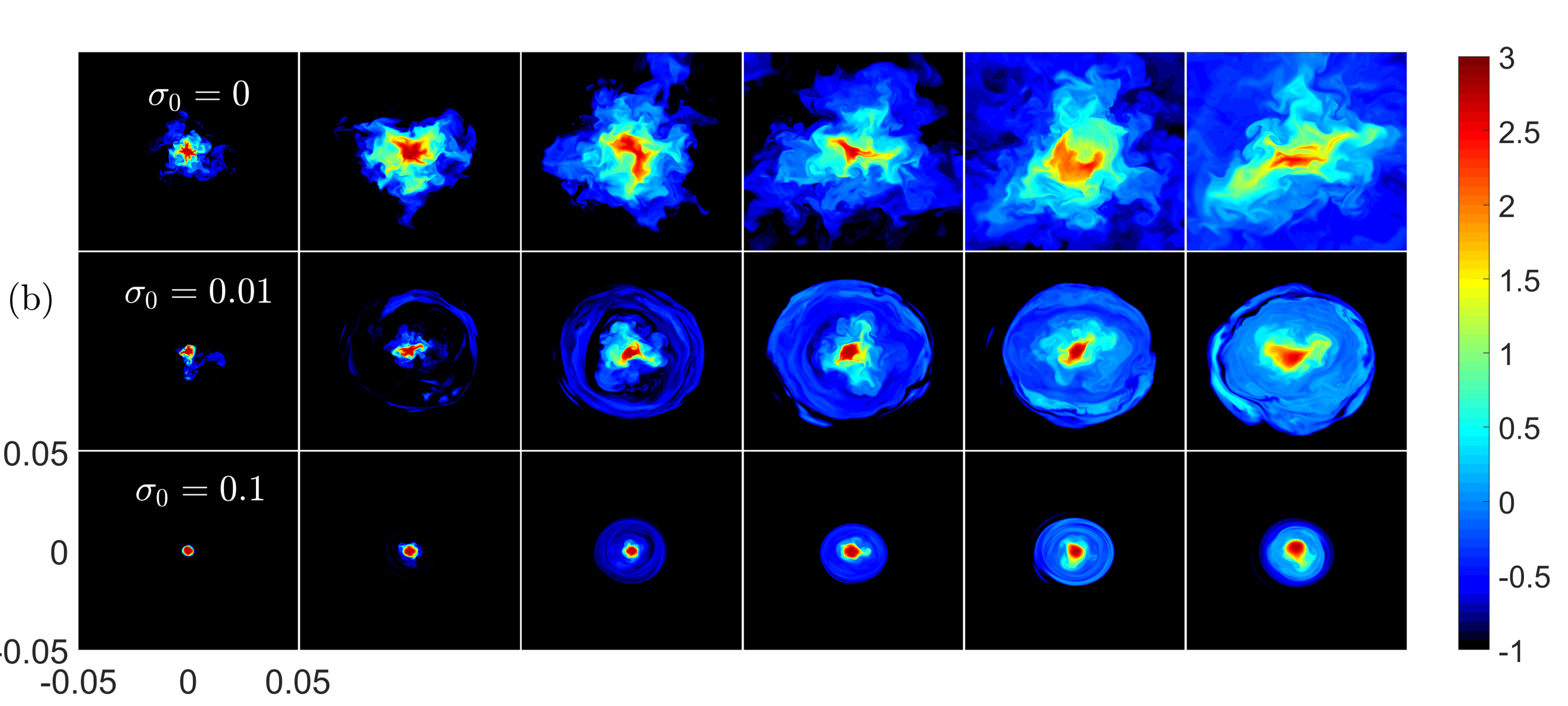}
		\includegraphics[scale=0.26,trim=6cm 0cm 0cm 0cm]{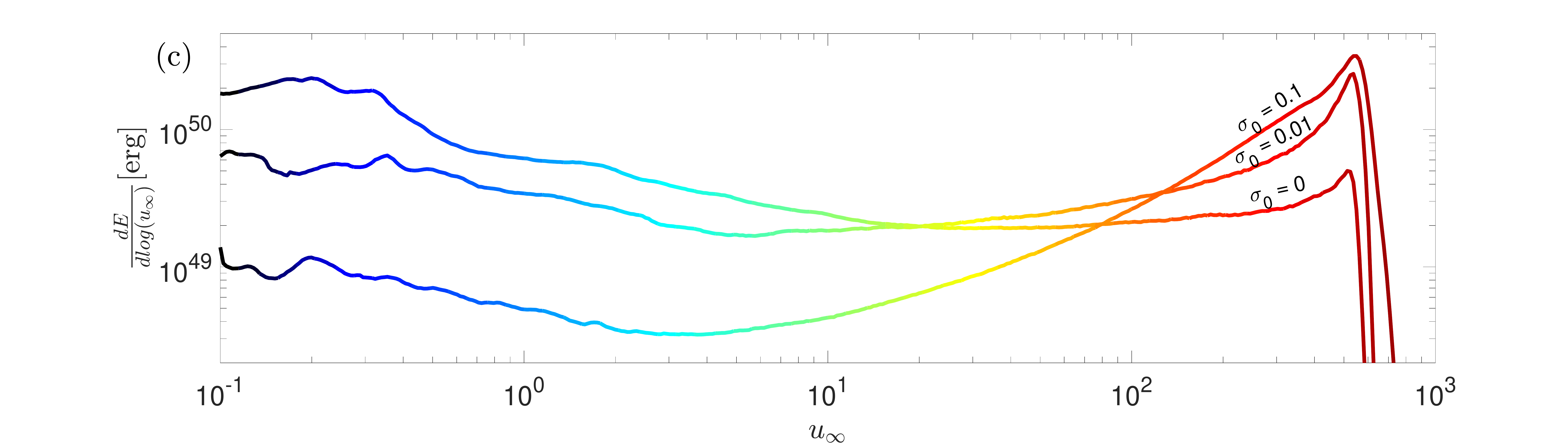}
		\caption[3D magnetized jet]{
			The terminal velocity $ u_\infty \equiv h\Gamma-1 $ if no further mixing takes place, in models $ \A, \C $ and $ \D $ upon breakout.
			(a) The log$ (u_\infty) $ on the $ \hat{x}-\hat{z} $ plane, parallel to the propagation axis of the jet.
			The dashed white lines mark the locations at which the $ \hat{x}-\hat{y} $ plane cuts (perpendicular to the propagation axis of the jet) in panels (b) are taken.
			(c) The energy distribution per logarithmic scale of the terminal four velocity. The colors show the mixing of the different elements, in correspondence to the terminal four velocity maps. Only material above the collimation shock, where mixing is present is considered.
			Videos of the evolution of the instabilities in the hydrodynamic and magnetic simulations are available at \url{http://www.astro.tau.ac.il/~ore/instabilities.html}.
		}
		\label{fig:magnetic_cuts}
	\end{figure*}
	
	The stabilization effect of a weak magnetic field on the structure of propagating jets is demonstrated in models $\A, \B, \C$, and $\D$. The models show the evolution of a lGRB jet propagating in a stellar envelope, where only the strength of the injected magnetic field varies. We find that a value of $ \sigma \gtrsim 10^{-2} $ is sufficient to inhibit the growth of instabilities on the jet boundary.
	Figure \ref{fig:magnetic_cuts} depicts the $\LogT$ of $u_{\infty}\equiv \Gamma h-1$, which is a measure for the degree of mixing in each fluid element, in the jets of 
	models $\A$, $\C$, and $\D$, close to their breakout from the stellar surface (the jet evolution and the mixing in model $ \B $ are similar to those seen in model $ \A $). The panels from top to bottom show: (a)  meridian cuts along the jet axis, (b) cross-sectional cuts at different altitudes, where the white dashed lines in panels (a) mark the locations of the cuts; and (c) the distribution of energy in the box (excluding the rest mass) in bins of $\LogT u_{\infty}$. We account for the energy in the region above the convergence point of the first collimation shock to exclude contributions from unshocked jet material. We distinguish between four domains, which correspond to different values of $u_\infty$. Fluid elements with $ u_\infty \gtrsim 100 $ are typically associated with the light jet material. Elements with $ 100\gtrsim u_\infty \gtrsim 3 $ correspond to the JCI, where jet material underwent mixing with the mildly-relativistic cocoon. Elements with $ 3 \gtrsim u_\infty \gtrsim 0.1 $ correspond to the inner cocoon, while elements with $ u_\infty \lesssim 0.1 $ correspond to the outer cocoon and to the unshocked stellar material. The outer cocoon, which is less relevant for this discussion, is not seen here due to a limited range of the color scale. It can be seen together with the other three zones in Appendix \ref{sec:outer_cocoon}, where we plot the mass density, $\sigma$ and $u_\infty$ on meridian slices at a wider color scale range.
	
	Figure \ref{fig:magnetic_cuts}a shows meridian cuts of the jets through the $ \hat{x}-\hat{z} $ plane.
	The pure hydrodynamic jet is much less stable than the magnetized jets. Mixing between jet and medium material takes place in three regions: the jet head, the interface between the inner and the outer cocoon and the interface between the inner cocoon and the jet (the JCI). The result of the first two is that the inner cocoon of shocked jet material, is strongly mixed with medium material.
	In pure hydrodynamic jets, the mixing in the third region along the JCI results in a diffused jet-cocoon structure which gradually penetrates into the jet core and eventually reduces the Lorentz factor in the entire jet.
	After the jet breaks out, the point where the instabilities in the JCI reach the jet axis and erode the jet, moves down until it reaches the point just above the convergence point of the first collimation shock.
	When adding a weak toroidal magnetic field with $ \sigma_0 = 10^{-2} $, the hydrodynamic instabilities on the JCI relax considerably. The inner cocoon is still heavily mixed (mostly through the interactions at the jet head), but the subdominant field stabilizes the jet interface enough, allowing the jet core to remain intact all the way to its head. The stabilization effect is also evident on the interface between the inner and outer cocoon.
	Increasing the magnetic field to $ \sigma_0 = 10^{-1} $ makes the jet even more stable, and it retains almost all its energy in an unmixed form.
	Extremely weak field with $ \sigma_0 = 10^{-4}$ have no effect on the jet, which behaves almost exactly as the hydrodynamic jet.
	
	The stabilization of the jet boundary by  magnetic fields in lGRBs leads to a faster propagation of the jet head through the star. This has an impact on the cocoon morphology and its energy content. The cocoon receives its energy from the jet head during the time the head propagates through the star. Once the head breaks out it accelerates to relativistic velocities and the energy injection into the cocoon stops. It follows that the energy in the cocoon holds  $E_c\simeq L_jt_{\rm b}(1-\bar{\beta}_{\rm h})$, 
	where $\bar{\beta}_{\rm h}$ is the average propagation 3-velocity (in units of $ c $) of the head through the star. The $(1-\bar{\beta}_{\rm h})$ term accounts for the decrease in energy flow into the head due to the relative motion of the head with respect to the relativistic jet material \citep{Bromberg2011b}.
	The faster head velocities seen in jets with higher $\sigma_0$ imply higher $\bar{\beta}_{\rm h}$ and smaller $t_{\rm b}$, altogether resulting in less energetic and narrower cocoons. This trend can clearly be seen in figure \ref{fig:magnetic_cuts}a.
	In our canonical configurations, jets with stronger magnetic fields ($ \sigma_0 \gtrsim 10^{-2} $) propagate with an average velocity of $ \bar{\beta}_{\rm h} \approx \frac{2}{3} $ inside the star, three times faster than hydrodynamic jets which propagate with an average velocity of $ \bar{\beta}_{\rm h} \approx \frac{2}{9} $.
	
	Figure \ref{fig:magnetic_cuts}b depicts cross sectional cuts of the jets on the $\hat{x}-\hat{y}$ plane at altitudes matching the white dashed lines at Figure \ref{fig:magnetic_cuts}a. The hydrodynamic jet features RTI fingers that begin to grow on the jet boundary at the collimation point. These are amplified by the RMI above the convergence point of the collimation shock at $z\simeq0.1 R_*$ (see \href{http://www.astro.tau.ac.il/~ore/instabilities.html}{movie}). The instabilities grow with $z$ and disrupt the jet core above $z=0.6 R_*$. When introducing magnetic fields the instabilities are weaker, resulting in smaller mixing of jet and cocoon material and allowing the jet to maintain cores with higher $ u_\infty $. For $\sigma_0=0.1$ the jet core is hardly affected by the instabilities and the mixing at the JCI is negligible.

	Figure \ref{fig:magnetic_cuts}c depicts the distribution of the total energy excluding rest mass in the box, $\int (T_{00}-\rho\Gamma) dV$, per logarithmic unit of $u_\infty$. The integration is taken above the first collimation shock to exclude unshocked jet material.
	The color scheme is the same as in Figure \ref{fig:magnetic_cuts}a,b: The jet is shown in red ($ u_\infty \gtrsim 100 $), the jet-cocoon interface in yellow-green ($ 100 \gtrsim u_\infty \gtrsim 3 $), the light to dark blue ($ 3 \gtrsim u_\infty \gtrsim 0.1 $) depicts the inner cocoon and the dark blue to black ($ u_\infty \lesssim 0.1 $) the outer cocoon.
	By the time the hydrodynamic jet reaches the stellar surface,  most of its energy is deposited in the cocoon, featuring a monotonic energy decrease towards higher terminal velocities, and only a small fraction of unmixed jet material is left in the jet. The JCI contains a comparable amount of energy to that of the jet and features a flat distribution of energy per logarithmic unit of $u_\infty$.
	Magnetic jets maintain a larger fraction of material in an unmixed state. It is manifested as a change in the trend of the JCI profile featuring an increase of energy with  
	$u_\infty$ towards a prominent peak at velocities associated with the unmixed jet material.
	
	\begin{figure}
		\centering
		\includegraphics[scale=0.17]{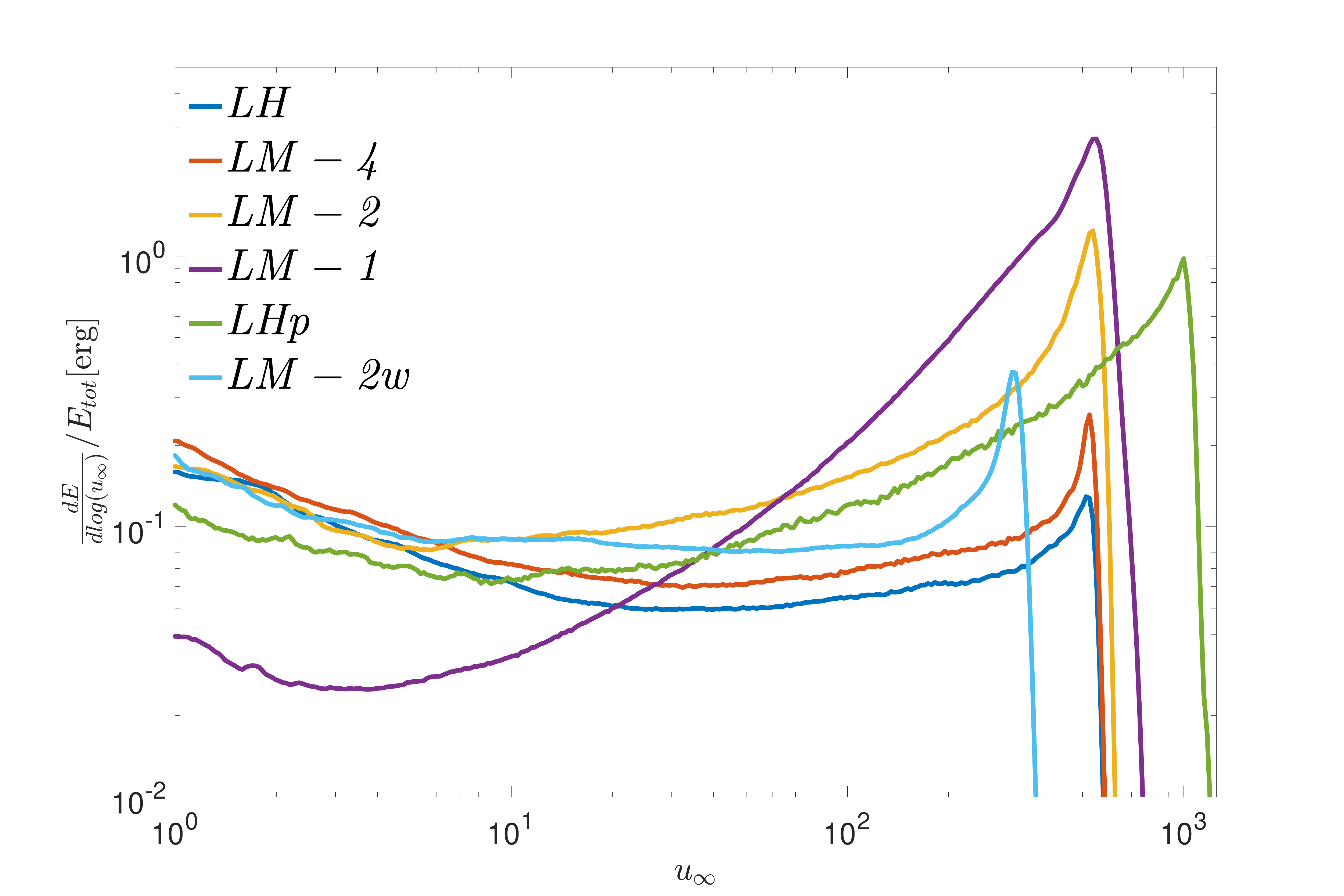}
		\caption[]{The energy distribution per a logarithmic scale of $ u_\infty $ for all lGRB models upon breakout. The energies are normalized by the total energy in each model. Note however that since the region $ u_\infty < 1 $ is not shown in the figure, the integrated energy in the figure is not the same for all models.
		}
		\label{fig:hg_plots}
	\end{figure}
	
	Figure \ref{fig:hg_plots} depicts the energy distribution per logarithmic unit of $u_\infty$ for all lGRB models, similar to Figure \ref{fig:magnetic_cuts}c.
	It shows a clear trend between models $ \A, \B, \C$, and $\D $, which differ from each other by the strength of the injected magnetic field, with stronger fields leading to faster and more stable jets.
	One effect of the increased stability is an inverse correlation between $\sigma_0$ and the breakout time (see Table \ref{table}).
	
	In our canonical setup we find that a magnetic field with $ \sigma_0 \sim 10^{-4}$ is not strong enough to have a significant stabilizing effect and only when the initial magnetic field is amplified to $ \sigma_0 \gtrsim 10^{-2} $, it becomes strong enough to stabilize the jet boundary. We caution that the minimal value of $ \sigma_0 $ that leads to jet stabilization depends on the properties of the system. We find that the two properties that have the most notable effect on the jet stability, besides the magnetization, are the jet injection angle, $\theta_{j,0}$ and the ratio between the jet energy and the medium density. Thus wider jets with lower luminosity and/or higher medium density require higher values of $\sigma_0$ to become stable.
	These dependencies are illustrated in models $ \E $ and $ \F $.
	Model $ \E $ features a hydrodynamic jet with half the opening angle and seven times more power than in our canonical model. The jet has a similar breakout time and $u_\infty$ energy distribution as those of the magnetized jet with $\sigma_0=10^{-2}$ in model $\C$. Note that here we discuss the jet structure before breakout. As we show below (\S\ref{sec:post_breakout}), after the breakout the evolution of the magnetized flow (model $\C$) is different than that of model $ \E $, due to the magnetic stabilizing effect.
	In model $ \F $ we consider a magnetized ($ \sigma_0 = 10^{-2} $) jet with a wider injection angle, for which the curved streamlines in the collimation shock undergo stronger mixing. This jet shows a similar behavior to the canonical hydrodynamic jet in model $ \A $, with a similar breakout time and a rather flat energy distribution. Namely, it is unstable and requires a stronger magnetic field to stabilize it.
	
\begin{figure}
		\centering
		\includegraphics[scale=0.23]{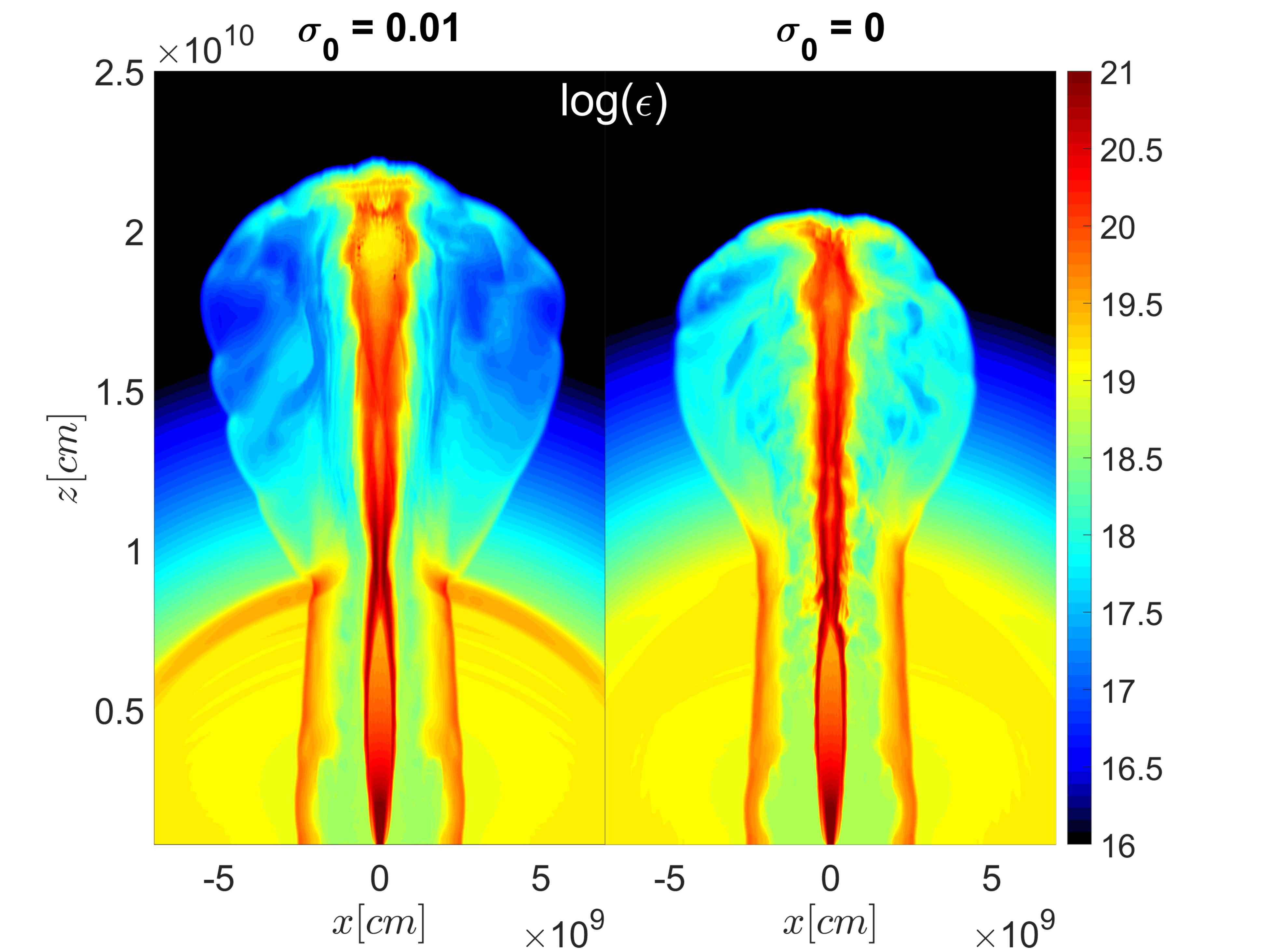}
		\caption[]{
			Logarithmic energy density $ \epsilon [\erg\cm^{-3}] $ map of magnetized (left) and hydrodynamic (right) sGRB jets after breakout from the core ejecta, one second after injection (1.6s after the merger). 
		}
		\label{fig:sgrb_e}
	\end{figure}

\begin{figure}
		\centering
		\includegraphics[scale=0.33]{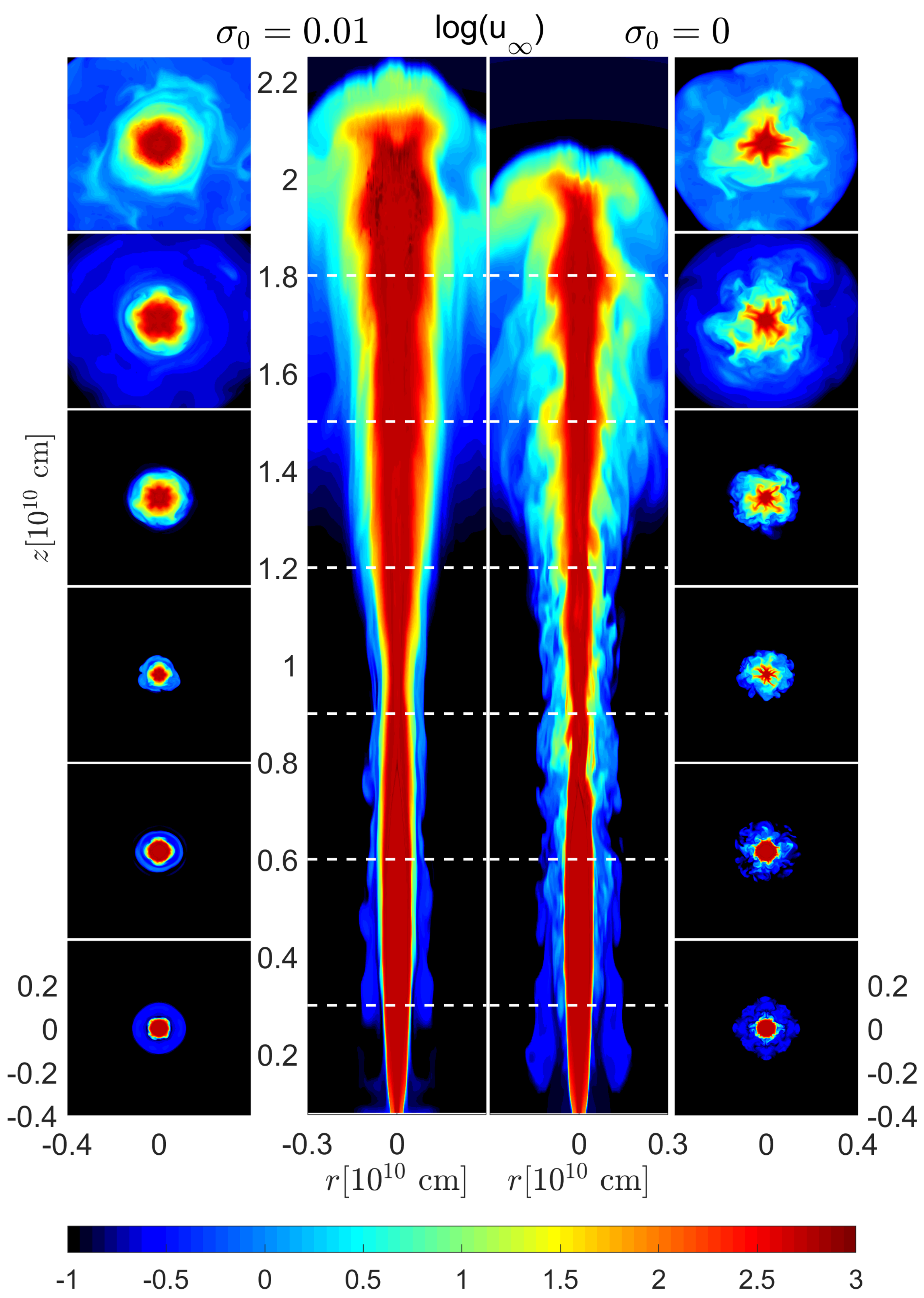}
		\caption[]{
			Logarithmic $ u_\infty $ maps of magnetized (left) and hydrodynamic (right) sGRB jets after breakout from the core ejecta. The time is the same as in Figure \ref{fig:sgrb_e}, one second after injection (1.6s after the merger).
			The middle panels depict an $ \hat{x}-\hat{z} $ view and the side panels show an $ \hat{x}-\hat{y} $ cuts, similar to Figure \ref{fig:magnetic_cuts}. The boundary of the magnetic jet is stable to a good approximation. The boundary of the hydrodynamic jet is somewhat less stable mostly above convergence point of the collimation shock, however the instability does not reach the jet core as in the case of lGRBs. 
		}
		\label{fig:sgrb_u}
	\end{figure}

	Hydrodynamic sGRB jets are more stable than hydrodynamic lGRB jets, owing to their lighter surrounding medium. Therefore, it is reasonable to expect that magnetic fields will not alter the jet behavior at the same extent as in lGRBs. Figure \ref{fig:sgrb_e} depicts the jets of models $ \G $ and $ \h $ after breakout from the core of the NS merger ejecta, one second after launching. It shows similarly stable jets propagating at almost the same velocity and having comparable structures. In Figure \ref{fig:sgrb_u}, which depicts cross sectional cuts of the four-velocity along the jet, it is seen that the hydrodynamic jet is somewhat less stable, developing RTI fingers, which become evident above the convergence point of the collimation shock, at $z \approx 10^{10}$ cm. 
	This result demonstrates that magnetic fields help in stabilizing sGRB jets as well. Yet, the combination of a lighter medium with a short breakout time inhibits a further growth of the instabilities in the hydrodynamic jet, so that the jet core remains intact and the global jet structure is similar to that of the magnetized jet.
	If the jet engine remains active for a sufficiently long time, the collimation shock will move out of the ejecta, open up and inhibit further RTI growth.
	In our sGRB models we find that $ \sim 2 $ seconds after the jet is launched, the collimation shock exits the ejecta and the baryon entrainment stops.
	
	\subsection{Conditions at the jet base}
	\label{sec:effect}
	
	The jet dynamics before the breakout is dictated to a large extent by the conditions at the jet base, where the collimation by the cocoon pressure takes place and the jet cross section is set. When examining the structure of magnetized lGRB jets in our simulations we identify a previously unnoticed structure that is formed at the base of the jet in some of our simulations, and has a significant effect on its propagation up to the breakout point. The same structure was absent from all the hydrodynamic simulations presented in GNB20, except for model $\E$. Below we discuss this structure and its effect.
	
\begin{figure}
		\centering
		\includegraphics[scale=0.25]{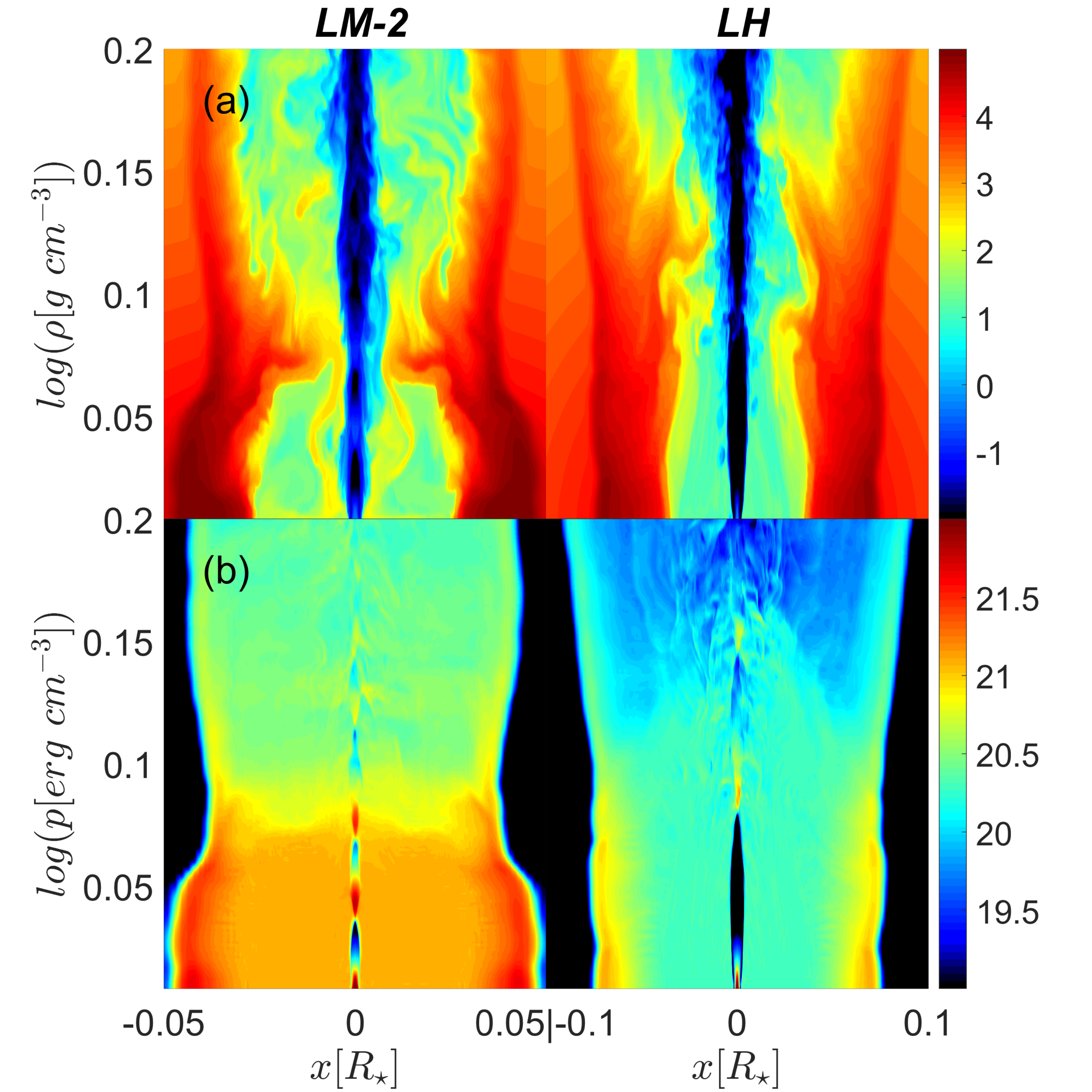}
		\includegraphics[scale=0.25]{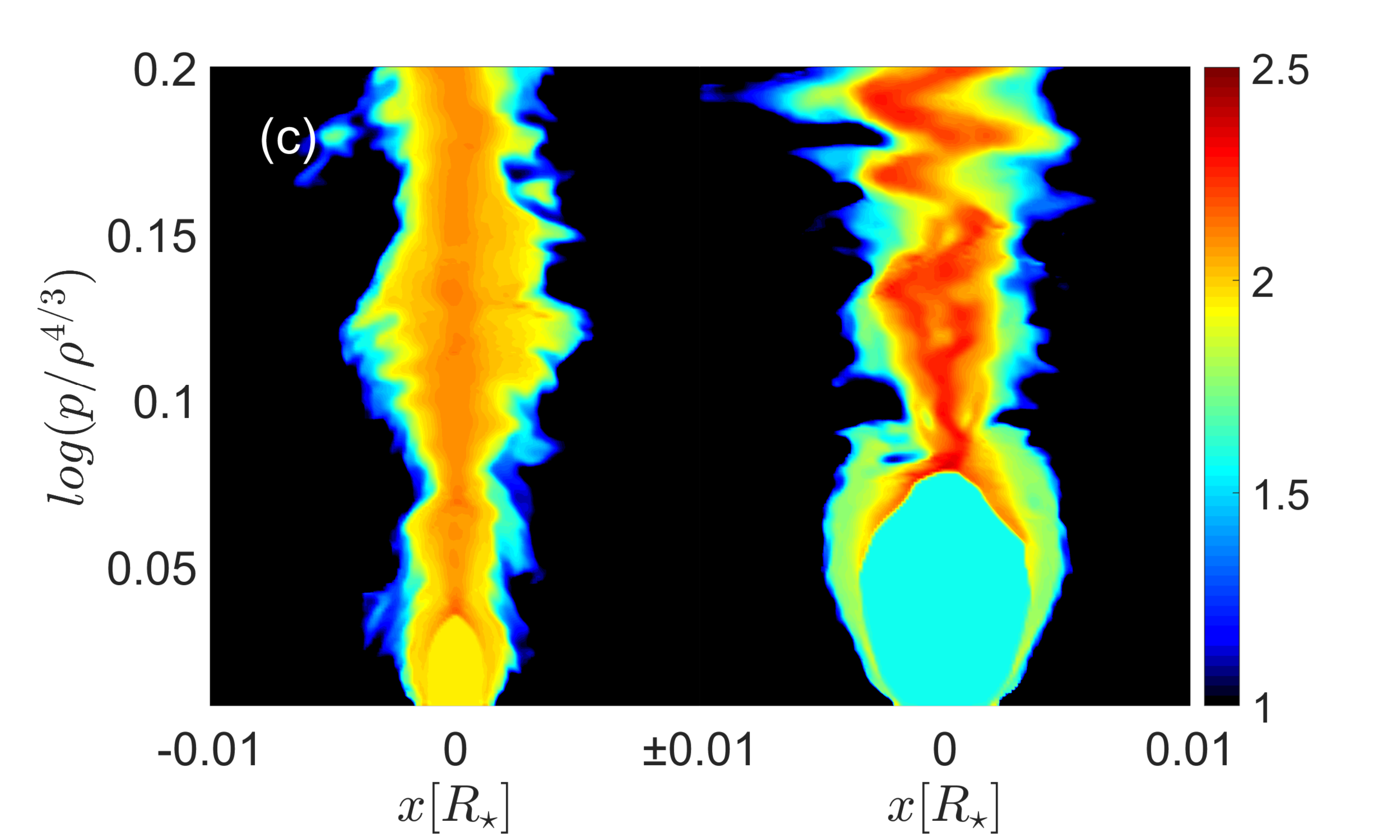}
		\includegraphics[scale=0.26]{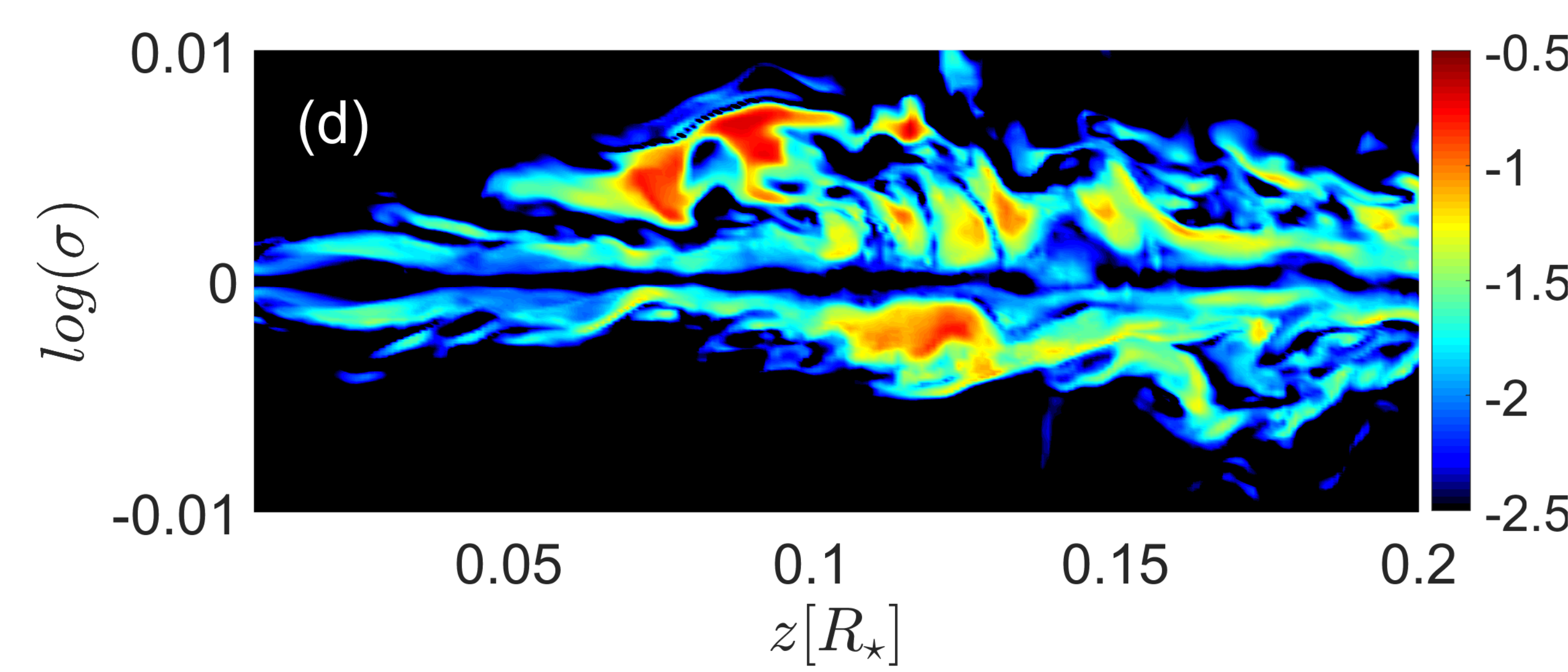}
		\caption[Pressure maps]
		{
			Cross sectional cuts of two jet bases in logarithmic scales for models $ \C $ (left, $ \sigma_0 = 0.01 $) and $ \A $ (right, $ \sigma = 0 $). (a) mass density ($ {\rm g}\cm^{-3}$), (b) pressure ($ \erg\cm^{-3} $), (c) entropy (d) $ \sigma $ of model $ \C $. The snapshots are taken just before the jets break out of the stars. 
			A ``pocket" of high pressure is evident at the base of model $ \C $. It contributes to the stabilization of the jet, leading to a narrower faster propagating jet. The difference in the widths of the jet cocoons in models $\A$ and $\C$ is the result of the shorter breakout time of model $\C$.
			Note: the x-scales of panels a+b ($\pm0.05$ magnetic, $\pm0.1$ hydrodynamic) capture the cocoon properties, while the x-scales of panels c+d ($\pm0.01$) resolve the jet interior.
		}
		\label{fig:feedback}
	\end{figure}

	Figure \ref{fig:feedback} shows the base of two representative lGRB jets: a magnetized jet (model $\C$, left panels) and a hydrodynamic jet (model $\A$, right panels). A region of high pressure is evident around the base of the magnetized jet (Figure \ref{fig:feedback}b), which is separated from the cocoon above it by a high density barrier (Figure \ref{fig:feedback}a). The barrier prevents pressure equilibration with the cocoon, preserving the high pressure region throughout the propagation of the jet though the star and for a similar amount of time after the breakout.
	We term this region as the {\it pocket}.  
	The high pressure at the pocket enhances the jet collimation at its base substantially. This appears to stabilize the jet from boundary instabilities, as fresh jet material passing through this structure does not develop significant RTI fingers. After the jet material exits the pocket it expands gradually, adjusting to the lower pressure at the cocoon. Due to this gradual expansion, the stable conditions are also maintained above the pocket and help keeping a low mixing level along the entire jet, up to its head, during the lifetime of the pocket.
	
	We witness the pocket formation and jet stabilization in both magnetic jets and high power, narrow hydrodynamic jets.
	A plausible explanation for the boundary stabilization may originate from a collection of several mechanisms that take place in the pocket and help stabilizing the jet:
	(i) The high pocket pressure (Figure \ref{fig:feedback}b) imposes a jet collimation deep in the pocket at a much lower altitude than under a regular cocoon pressure. The smaller altitude implies that an instability has less time to grow in a fluid element before it reaches the convergence point of the shock. 
	(ii) The convergence point of the first collimation shock is important for the growth of boundary instabilities. Above it the converging jet flow hits the axis and forms a reflection shock, which drives impulsive RMI that substantially increases the amplitude of the RTI fingers. When a pocket is present it typically contains the convergence points of the first and the second collimation shocks. The high pressure and subsequently high temperature\footnote{In a relativistic equation of state the temperature is proportional to $p^{0.25}$.} imposed on the jet flow by the pocket results in weaker reflection shocks, which in turn leads to weaker RMI and a smaller amplification of the already diminished RTI modes. The difference in the reflection shocks strength in jets with and without pockets can be seen in Figure \ref{fig:feedback}c.
	(iii) A second effect of weaker reflection shocks is that the jump conditions across the shocks are milder, thus the jet fluid that crosses the reflection shocks suffers smaller decrease in its Lorentz factor and is able to maintain a faster steadier flow along the jet. 
	Since the instabilities evolve in the fluid frame, their evolution in the star frame will be slower due to time dilation. 
	(iv) Finally, in magnetized jets the magnetization, amplified by the passage through the collimation shock, may contribute to the stabilization as well (Figure \ref{fig:feedback}d).
	Conditions (i)-(iii) stabilize also hydrodynamic jets with pockets, while condition (iv) adds an extra stability to magnetized jets, making them even less susceptible to baryon contamination.
	The transition of the jet flow from the pocket to the cocoon region is not accompanied by a strong lateral expansion,
	thus the stable conditions obtained at the pocket are largely maintained also in the cocoon region. The sum of these effects leads to a relatively stable and unmixed jet which propagates fast through the star until it breaks out of the medium. After breakout the stabilization effect of the pocket slowly fades away. The reason is a fast decline in the pocket density and pressure following the breakout, which brings the conditions in the pocket close to those of a regular cocoon.
	
	The formation of the pocket is linked with a slow propagation of the jet head in the medium surrounding the jet base soon after it is first launched. If the jet head propagates with a velocity slower than the cocoon expansion velocity $v_{\rm c}\simeq\sqrt{p_{\rm c}/\rho_{\rm a}}$, where $p_{\rm c}$ and $\rho_{\rm a}$ are the cocoon pressure and the medium density respectively, the cocoon it generates expands quasi-isotropically, forming a quasi-spherical, highly pressurized region around the jet. As the jet head advances to regions with lower density, its velocity gradually increases until it overtakes the cocoon. At this stage the jet exits the pocket and forms a regular cocoon above it with a typically much lower pressure. In some cases, when the exit from the pocket occurs over a short time with respect to the reaction time of the medium surrounding the pocket, a high density barrier remains between the pocket and the cocoon (Figure \ref{fig:feedback}a), which prevents a pressure equilibration and prolongs the lifetime of the pocket. 
	
	In our simulations the pocket is formed soon after the jet is launched and it is affected by the specific initial and boundary conditions of the simulations (e.g., $z_0)$. In reality, the conditions that can lead to the formation of a pocket take place deep in stellar cores of lGRB progenitors, close to the location where the jet is launched. Convergence tests we conducted indicate that under the conditions tested here, the formation of the pocket is unavoidable. Nevertheless, the pocket formation and its survival over the long times seen in our simulations may be an artifact, which is affected by the numerical details of our simulations (such as the injection altitude, the boundary conditions etc.). As the pocket facilitates the jet stability, it implies that the distributions and mixing levels that we obtain for jets before the breakout may change. However, as we show in the next section, following the breakout the pocket is dispersed and jets with and without pockets are converging to have similar profiles (e.g., models $\A$ and $\E$). The result is that the pocket, if exists, has only a limited effect on the prompt emission and almost no effect on the afterglow emission.

	\section{Post-breakout evolution}
	\label{sec:post_breakout}
	
	As the jet-cocoon system breaks out from the dense medium, it expands and accelerates. The first signal that the jet produces is the prompt emission. At later times, the interaction with the circum-burst material drives a bow shock which generates the so-called afterglow emission. The afterglow signal depends on the mass distribution in the circum-burst medium and on the energy distribution of the jet-cocoon outflow. As we show below, the post-breakout energy distribution is modified by the presence of magnetic fields, leading to specific emission properties that may help identifying the presence of magnetic fields in GRB jets.
	
	To properly model the energy distribution in the outflow, one needs to follow the system up to the homologous phase, which takes place beyond several stellar radii. Modeling this with MHD simulations requires high resolution over a wide range of dynamical scales, which is too demanding for our current computational power. As a result, we cannot directly simulate the system evolution outside the star beyond two stellar radii.
	To overcome this problem we characterize the properties of the jet and the cocoon material 
	close to the stellar edge\footnote{We set the boundary of the simulation box at the stellar edge in the lGRB models and at a similar radius in the sGRB models (see table \ref{table}), and continue the simulations after the jet breaks out. We then characterize the outflow before it exits the medium and before it escapes the simulation boundary.}, and extrapolate them to the radii from which the emission is expected to originate. 
	In our hydrodynamic jets, where modeling the full dynamical range of the system is possible, we find that the baryon loading of elements is not affected by mixing outside the star, and effectively it freezes out once it exits the star (\citealt{Gottlieb2019b}; GNB20). The final four-velocity of each fluid element is therefore $ u_{\infty}=\Gamma h-1$, where $\Gamma$ and  $h$ can be measured at any radius outside of the star.
	Since magnetic fields inhibit the growth of hydrodynamic instabilities, thus weakening the strength of the jet-cocoon interaction, it is reasonable to assume that this property is maintained at an even larger accuracy in magnetized jets. Therefore, in our weakly magnetized simulations we assume that  elements retain their baryon loading after exiting the star.
	
	\begin{figure}
		\centering
		\includegraphics[scale=0.23]{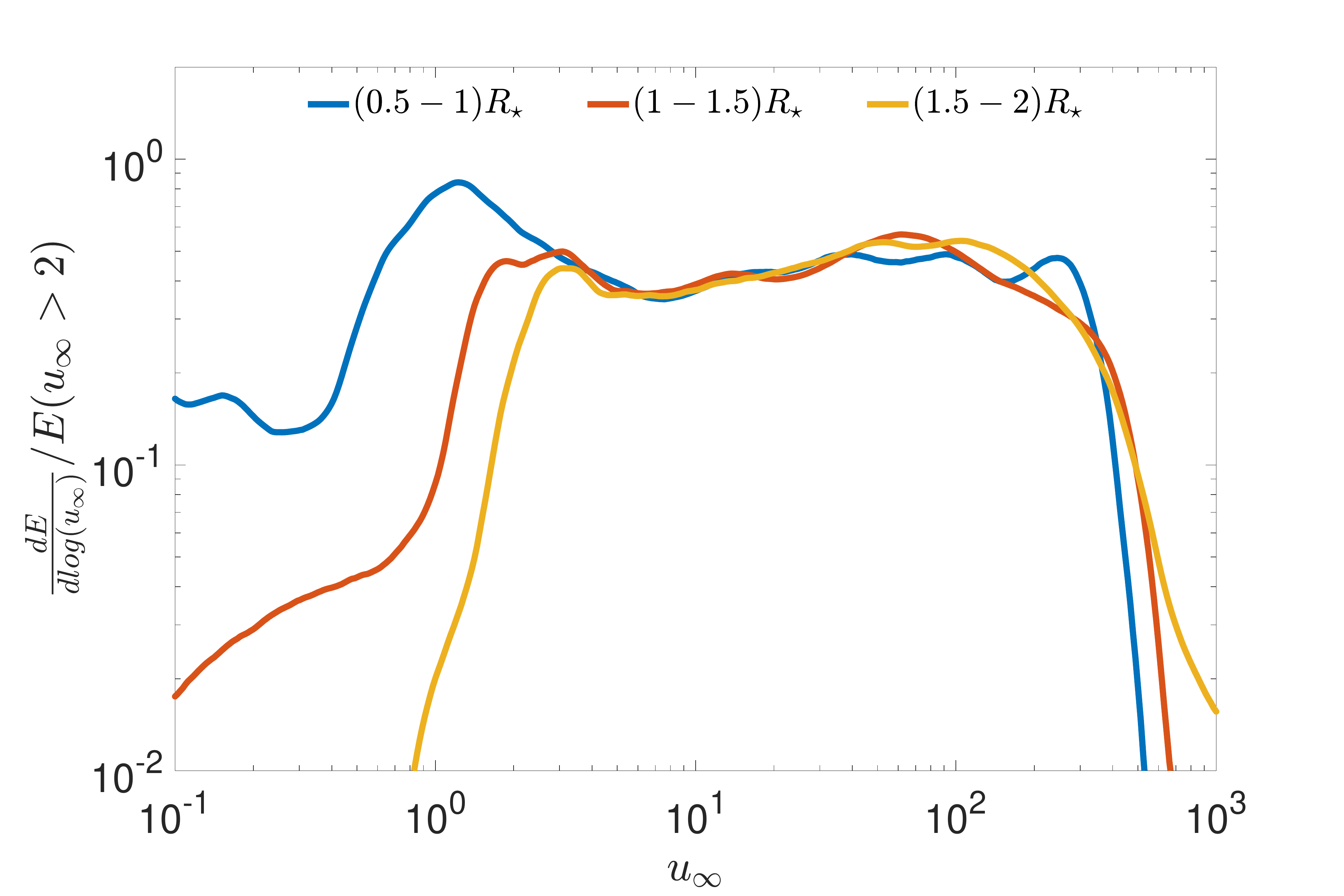}
		\caption[]{
			The evolution of the energy distribution in simulation $ \Ce $ of a $ 0.5R_\star $ sized slab of matter at different locations: when the slab is between $ z = 0.5R_\star $ and $ z = R_\star $ (blue), at $ R_\star < z < 1.5R_\star $ (red) and $ 1.5R_\star < z < 2R_\star $ (yellow).
			Our calculation assumes the slab to move at the speed of light, and hence does not properly account for the motion of the slab at $ u_\infty \lesssim 2 $.
			At $ u_\infty \gtrsim u_{\infty,\mx} $, numerical artifacts take place after breakout, and hence the curves disagree with each other.
			The energy is normalized by the total amount of energy at $ u_\infty > 2 $ at each time.
		}
		\label{fig:tracking}
	\end{figure}

	We verify this approximation by simulating the jet of model $\C$ in two ways. First we set the grid boundary at the stellar edge (at $R_*=10^{11}$ cm) and simulate the propagation of the jet through the star until breakout. We continue the simulation after breakout until the jet head reaches $4R_*$ (the jet head at this time is outside the simulation boundaries), and monitor the composition of the jet material in the simulated zone (at $r<R_*$). We then extend the simulation box of model $ \C $ to $2R_*$ and restart the simulation from the point where the jet is about to breakout (simulation $ \Ce $). We run the simulation until the jet head reaches $2R_*$ and verify that the composition of the jet-cocoon material at $r<R_*$ is identical in the two cases. We examine the composition of the material that breaks out of the star in the second simulation and verified that it does not change outside the star.
	In Figure \ref{fig:tracking} we present a tracking of a given slab of matter from inside the star up to two stellar radii. The blue line depicts the distribution of matter between $ z = 0.5R_\star $ and $ z = R_\star $ upon breakout. The red and yellow lines follow this slab of matter at different heights outside the star, where it is assumed to move at the speed of light.
	It is shown that the energy distribution at the JCI and in the jet remains unchanged after breakout in the range of $2\lesssim u_\infty\lesssim500$. At the low end of the distribution the curves differ from each other, since elements with  $ u_\infty \lesssim 2 $ move much slower than $c$,  and are therefore not tracked well. During the propagation outside of the star the simulation ($ \Ce $) develop numerical artifacts that appear in several individual cells outside of the star after the jet breaks out. The cells contain negligible amount of energy and their effect on the jet is small when the jet head reaches $2R_*$, therefore we could safely follow the jet head up to that radius. The effect of the numerical artifacts is seen as a small increase in energy in the yellow curve at $u_\infty > 500$. It can be seen that the contribution from these pixels is negligible, with a total energy of $<1\%$ out of the total energy of the outflow, and thus it influences neither the jet dynamics nor its properties.
	
	\subsection{Energy distribution}
	\label{sec:distributions}
	
\begin{figure}
		\centering
		\includegraphics[scale=0.23]{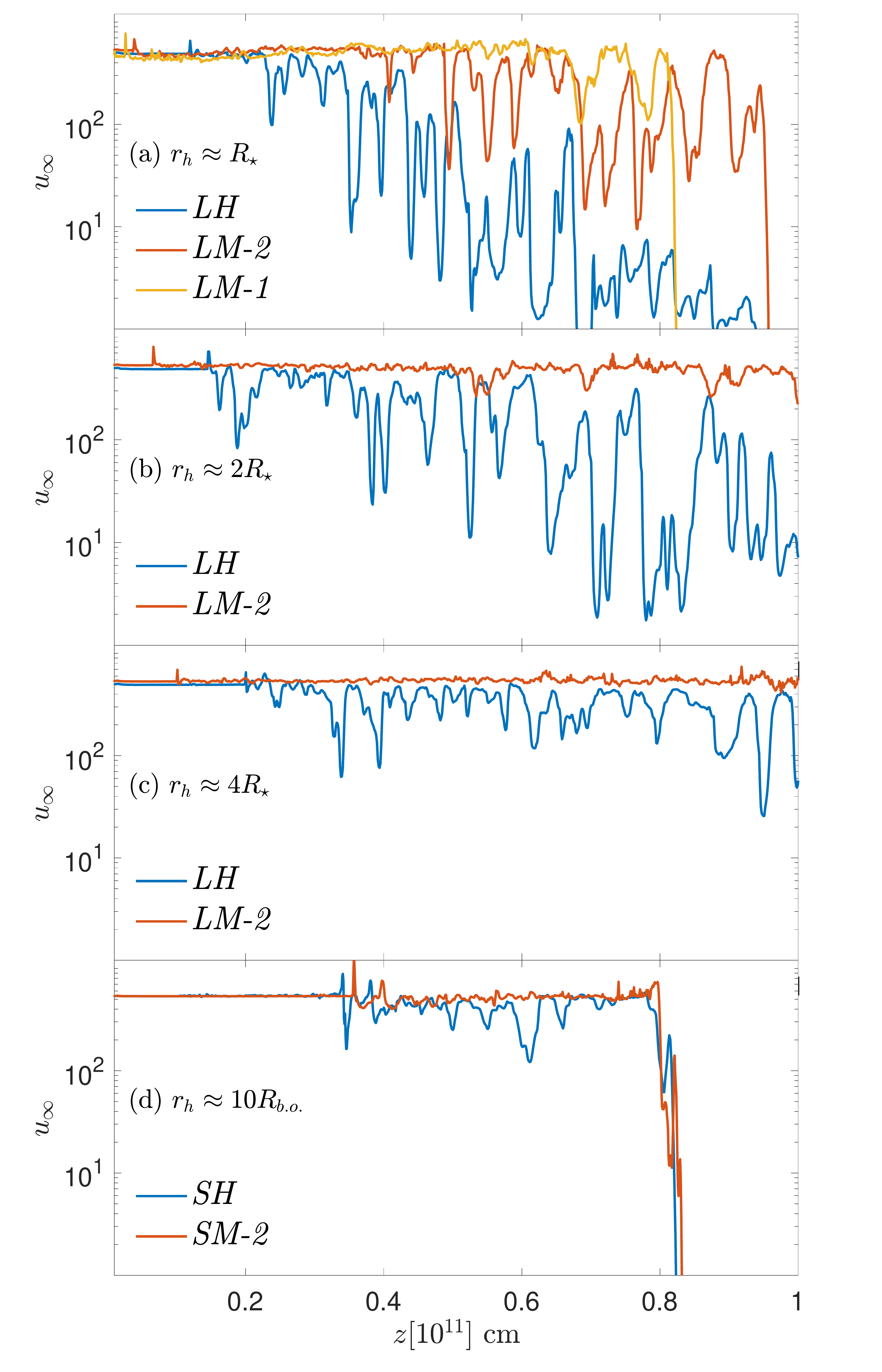}
		\caption[]{
			The temporal evolution of the mixing level at the  jet core, manifested by the quantity $ u_\infty $ on the jet axis. The location of the jet head, $ r_h $, at the time that the profiles are taken is shown.
			In panels (a,b,c) the evolution of different lGRB models is shown when the jet head reaches:
			(a) the stellar edge, (b) two stellar radii, and (c) four stellar radii.
			Panel (d) depicts the profile of the sGRB jets after breakout when the jet head reaches $ \sim 10 $ breakout radii.
		}
		\label{fig:hg_axis}
	\end{figure}

	In Figure \ref{fig:hg_axis} we demonstrate the effect of the mixing at the jet boundary on the baryon loading in the jet core before and after the jet head breaks out from the dense medium.
	Panels (a,b,c) depict the profile of $ u_\infty $ along the jet axis inside the star ($r<R_*$) for different models. The panels show the profiles at different stages: (a) before breakout, (b) when the jet head is at two stellar radii and (c) when the jet head is at four stellar radii.
	Panel (d) depicts the profile of $ u_\infty $ on the jet axis for the two sGRB models at late times.
	Unmixed jet material maintains its initial $ u_{\infty,0} = h_0\Gamma_0 $, whereas mixing reduces the value of $ u_\infty $.
	Prior to breakout, the hydrodynamic jet (blue curve) becomes unstable as soon as the second recollimation shock converges to the axis at $z=0.25R_\star$. As a result the material along the jet axis becomes highly mixed, featuring $ u_\infty < 100 $ at $ z > 0.5R_\star$. For $ \sigma_0 = 10^{-1} $ (yellow line) the initial $ u_\infty $ is by large conserved on the axis, whereas for $ \sigma_0 = 10^{-2} $ (red line) the jet is subject to a mild mass entrainment from $ z \approx 0.5R_{\star} $.
	After breakout the instabilities inside the star somewhat relax and the jet material becomes less loaded with baryons, featuring higher values of $ u_\infty $ \footnote{The simulation of model $\D$ with $\sigma=0.1$ crashes soon after the jet breakout, thus we could only show its pre-breakout values of axial $u_\infty$.
		Since this model features a stable jet at early times and magnetized jets become more stable as time progresses, it is expected to be at least as stable as model $ \C $ soon after breakout.}.
	However, while the weakly magnetized jet ($ \sigma = 10^{-2} $) has a quasi-flat distribution with almost no mixing inside the star, the hydrodynamic jet is still subject to a significant baryon loading, and shows fluctuations above the first collimation shock.
	
	\begin{figure*}
		\centering
		\includegraphics[scale=0.23]{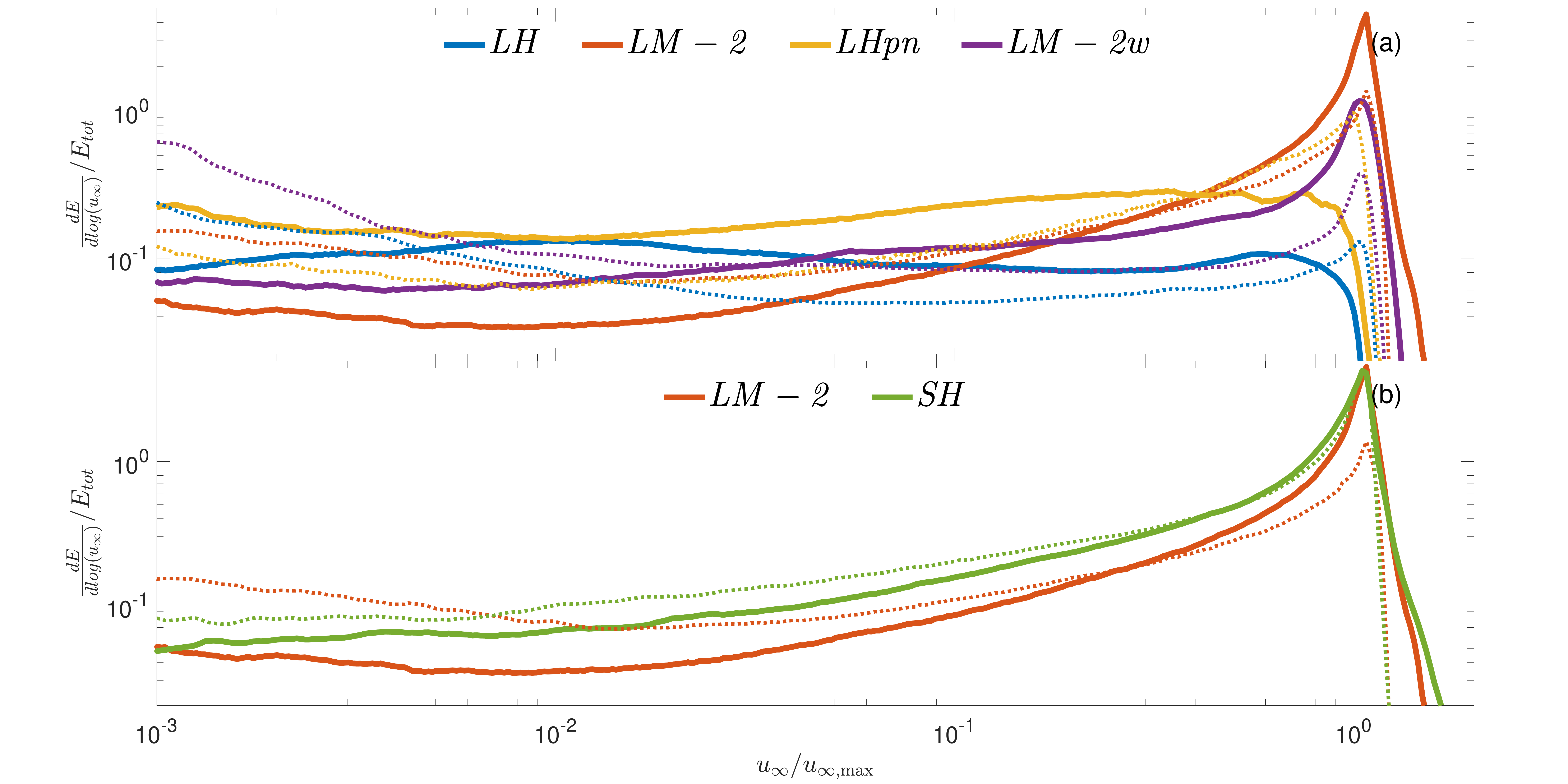}
		\caption[]{
			The energy distribution above the collimation shock per a logarithmic scale of $ u_\infty $. The $ \hat{x} $-axis of each model is normalized by its $ u_{\infty,\mx} $.
			The dotted lines reflect the distribution inside the dense medium upon jet breakout. The thick solid lines represent the distribution inside the dense medium when the jet head reaches four breakout radii.
		}
		\label{fig:hg_evolution}
	\end{figure*}

	In Figure \ref{fig:hg_axis}d we show the axial distribution of $u_\infty$ in our sGRB models after the jets broke out out from the ejecta, 3.6s after the merger.
	Only minimal baryon contamination reaches the spine of both jets, with the hydrodynamic jet (blue line) shows some rather minor fluctuations on the axis at $z \gtrsim 3.5\times10^{10}$ cm, above the convergence point of the collimation shock.
	In our sGRB simulations the collimation shock breaks out from the core ejecta $ \sim 2 $s after the merger (see Figure \ref{fig:sgrb_e}), and subsequently a further growth of the mixing is inhibited. As a result, the sGRB jets remain intact, showing only a limited evolution in time.
	This is in contrast to lGRB jets, which show evolution in time, however different between hydrodynamic and magnetized jets.
	
	In \S\ref{sec:mixing} we found that while the jet propagates inside the star, the stability of the jet boundary is primarily controlled by three parameters: {\it the jet power, the injection angle} and {\it the injected magnetic field}.
	Note that the (un)stabilizing parameters can cancel each other out. For example, magnetized jets with a large opening angle (model $ \F $) might still become unstable.
	In GNB20 we explored the two hydrodynamic effects. We found that all hydrodynamic lGRB jet models converge to a similar, roughly flat energy distribution in the four-velocity space long enough after the jet breakout.
	An example can be seen in Figure \ref{fig:hg_evolution}a, where we show the integrated values of $dE/dlog(u_\infty)$ up to $R_*$, at the time of the jet breakout (dotted lines) and when the jet head reaches four breakout radii (thick solid lines).
	It is shown that the hydrodynamic lGRB models $ \A $ (blue) and $ \E $ (yellow), which show very different distributions at the time of the breakout, exhibit quasi-flat distributions at later times. This behavior seems to be independent of the stability properties of the jets inside the star. Thus, the asymptotic distributions of all the hydrodynamic jets have a similar amount of energy in each logarithmic bin of the four-velocity. Namely, jets that were unstable and highly mixed when propagating inside the star become less mixed after breakout, while jets that were stable inside the star show the opposite evolution. The inverse evolution we observe in model $\E$ is likely the result of the dissolving pocket at the jet base, which takes place after the breakout (see \S\ref{sec:effect}).
	In sGRBs, the energy distribution in the four-velocity space inside the ejecta has a distinct peak at high $u_\infty$ which marks the relativistic jet, implying that the jet retains most of its energy, as shown for model $ \G $ (green) in Figure \ref{fig:hg_evolution}b.
	
	The temporal evolution of hydrodynamic jets inside the dense medium can be compared with that of magnetized lGRB jets (Figure \ref{fig:hg_evolution}a).
	Upon breakout the powerful hydrodynamic jet ($ \E $) shows a similar distribution to the canonical magnetized jet ($ \C $). Both are more stable than the wide angle magnetized jet ($ \F $) and the canonical hydrodynamic jet ($ \A $).
	When the jet heads reach $ 4R_\star $ (solid lines) 
	the distribution in both hydrodynamic jets flattens out. The distribution of the magnetized jets shows a distinct peak at $u_{\infty,\rm{max}}$ indicating that after breakout the core of magnetized jets retains most of its energy, even if it was initially less stable, e.g. in model $ \F $.
	
	The reason for what seems as a universal behavior, which depends only on the magnetization level, is that after the breakout all the jets open up due to a drop in the cocoon pressure, featuring common geometries. At this stage the interactions at the jet boundary remain strong enough to allow boundary instabilities to evolve. Thus,
	without magnetic fields the instabilities are able to erode lGRB jets and transfer a considerable amount of jet energy to the JCI, while magnetic fields prevent that and keep the JCI energy at a low level.
	Thus, while before the breakout the lGRB jet composition is governed by the three parameters discussed above, after breakout the effects of the initial jet opening angle and the luminosity to density ratio diminish and the mixing is almost exclusively determined by the injected magnetic field. In sGRBs the degree of mixing is substantially lower at all times in the hydrodynamic jet, and therefore, while the magnetized jet is slightly more stable, the difference between the two is minor.
	
	\subsection{Angular distributions}	
	
	\begin{figure}
		\centering
		\includegraphics[scale=0.23]{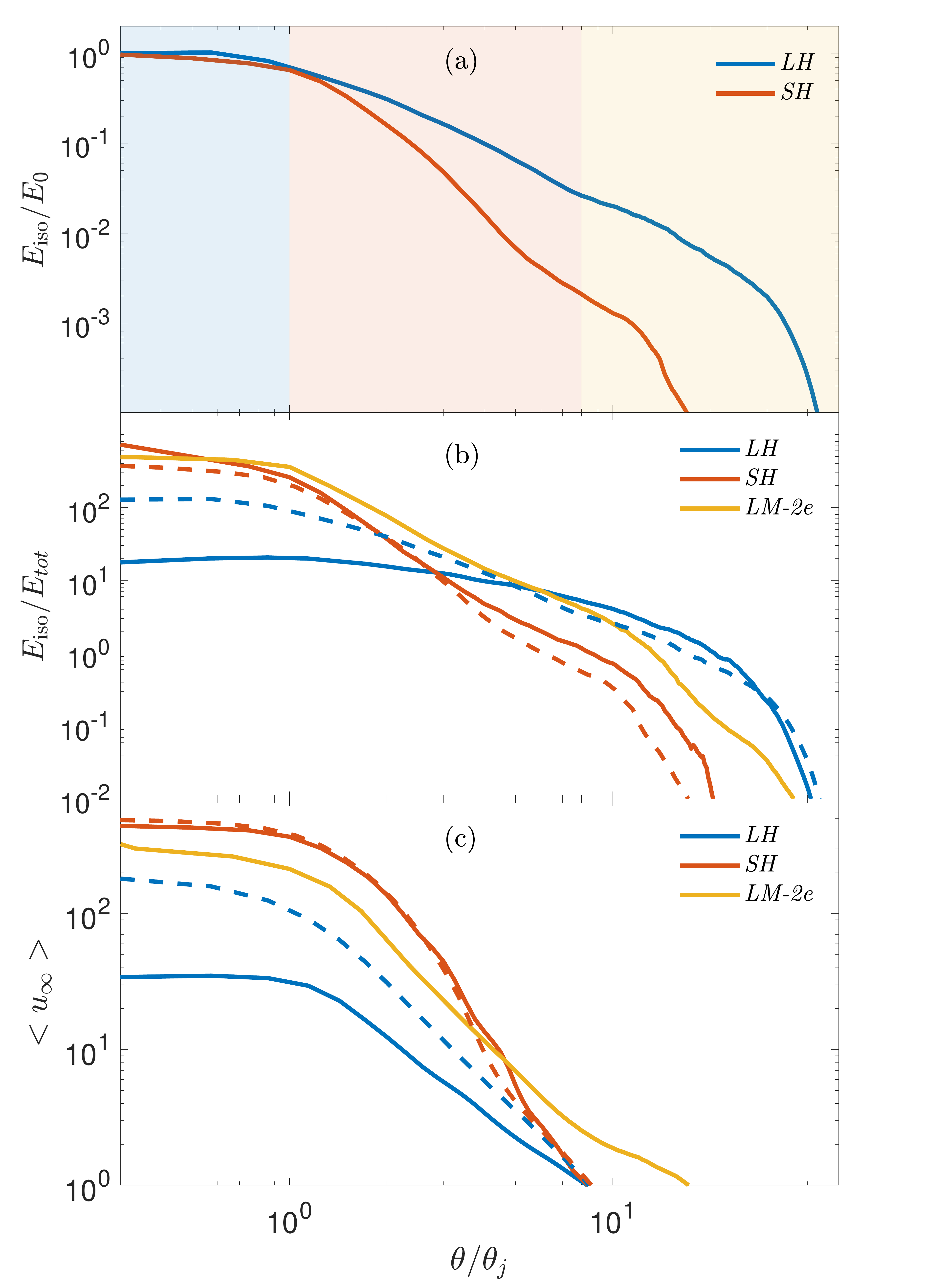}
		\caption[]{
			The angular distribution of all the matter that broke out of the dense medium. Shown when the jet head reaches 2 breakout radii (solid) and 10 breakout radii (dashed), after which the distributions do not change substantially.
			(a) The isotropic equivalent energy of the canonical hydrodynamic lGRB model $ \A $ and the hydrodynamic sGRB model $ \G $. The normalization is set by their energy at the core $ E_0 $, so that they coincide at the jet core. To ease the differentiation between the different regions, the background color is painted in pale blue for the jet region, in pale red for JCI and pale yellow for the cocoon.
			(b) The isotropic equivalent energy of the canonical hydrodynamic lGRB model $ \A $, canonical magnetized lGRB model, $ \Ce $ and hydrodynamic sGRB model $ \G $. The normalization is set by the total energy of each model.
			(c) The energy-weighted average four-velocity $ <u_\infty> $ along a line of sight of the canonical hydrodynamic lGRB model $ \A $, canonical magnetized lGRB model, $ \Ce $ and hydrodynamic sGRB model $ \G $.
		}
		\label{fig:hydro_structure}
	\end{figure}
	
	The asymptotic angular distributions of the outflow energy and velocity are of a particular interest as they dictate the late afterglow emission.
	In GNB20 we follow a variety of hydrodynamic jets after breakout and find that in all our simulations the jets feature common distributions composed of the following three components (see Figure \ref{fig:hydro_structure}a for the plot of isotropic equivalent energy when the jet head is at  10 breakout radii): 
	(i) A jet core (sometimes mixed with some cocoon material) having flat angular distributions of isotropic equivalent energy and four-velocity, extending up to $\theta_j \approx 0.2-0.4 \theta_{j,0}$ (pale blue background color).
	(ii) The JCI, extending from the edge of the jet core and up to $ \theta_c \sim 5-10 \theta_{j,0}$ (pale red background color). The angular distributions of the isotropic equivalent energy and of the four-velocity in the JCI can be fitted by power-laws with indices $ \delta $ and $ p_u $, respectively.
	(iii) The cocoon resides at larger angles ($ \theta > \theta_c $), showing a sharp cut-off in the angular distribution of the isotropic equivalent energy at an angle of $\sim 0.5-1 {\rm~rad}$. The angular energy distribution in this region can be fitted by an exponential cutoff, while the four-velocity distribution continues the power-law of the JCI with index $ p_u $ (pale yellow background color)\footnote{Note that the jet core is measured in units of $\theta_{j,0} $, whereas the typical angle of the cocoon is given in absolute values (see details in GNB20)}.
	Overall the angular isotropic equivalent energy distribution is described by:
	\begin{equation}\label{eq:Etheta}
	\Eiso\approx
	\begin{cases}
	\lambda E_0       & { \theta < \theta_j }\\
	\lambda E_0 (\theta/\theta_j)^{-\delta}  & { \theta_j < \theta < \theta_c }\\
	\lambda E_0 (\theta_c/\theta_j)^{-\delta}e^{-f_c(\theta-\theta_c)} & { \theta > \theta_c } \ ,
	\end{cases}
	\end{equation}
	where $ E_0 \equiv \frac{L_j(t-t_b)}{1-{\rm cos}\theta_j}$, $ t_b $ is the breakout time and $ \lambda $ is the fraction from the total injected energy after breakout, which is deposited into the jet core, $ \lambda \equiv E_j/L_j(t-t_b)$.
	The angular energy-weighted four-velocity distribution is described by:
	\begin{equation}\label{eq:utheta}
	u_\infty(\theta/\theta_j) =
	\begin{cases}
	<u_{\infty,j}>   & { \theta/\theta_j < 1 }\\
	<u_{\infty,j}> (\theta/\theta_j)^{-p_u}   & { \theta/\theta_j > 1 } \ .
	\end{cases}
	\end{equation}
	The value of $ <u_{\infty,j}> $ depends on the baryon loading in the jet core, where clean cores reach $ <u_{\infty,j}> \approx u_{\infty,\mx} $. Fitting this value to hydrodynamic models shows that $ <u_{\infty,j}> \approx 2\lambda u_{\infty,\mx} $.
	
	The hydrodynamic jets span a wide range of mixing degrees.
	In lGRBs, only a fraction $ \lambda \approx $ 0.2 of the original jet energy remains in the jet core. In addition, the power-law index of the angular energy distribution in the JCI is $ \delta \approx 2 $ (blue line in Figure \ref{fig:hydro_structure}a), owing to intense mixing. Short GRBs have the least mixing as they retain about half of their injected energy, $\lambda \approx 0.4$, and thus feature steeper angular energy distributions at their JCIs, with a power-law index $ \delta \approx 3.5 $ (red line in Figure \ref{fig:hydro_structure}a).
	Similarly, systems with less mixing such as sGRBs feature steeper four-velocity power-laws outside the jet core with an index $ p_u \approx 3 $ compared with $ p_u \approx 2 $ in hydrodynamic lGRBs.
	
	Performing a similar analysis to obtain the terminal angular distributions of magnetized jets requires following the outflow to a few stellar radii, which is beyond our current computational ability. Instead, we measure the distributions of the magnetized jet in simulation $\Ce$ at distances $<2R_*$, where we can directly simulate the system. These distributions provide an approximation of the final homologous distributions. We estimate the quality of this approximation by comparing the distributions in the hydrodynamic jets at $2R_*$ and $10R_*$.
	
	The angular distributions of the outflow outside the medium are shown in Figure \ref{fig:hydro_structure}. Panels (b),(c) depict the isotropic equivalent energy and four-velocity, respectively, in models $ \A, \G $ and $ \Ce $, showing only material that broke out from the media when the jet head reaches 2 breakout radii (solid lines) and 10 breakout radii (only for hydrodynamic models, dashed lines).
	Remarkably, we find that although the mixing and evolution of magnetized jets are different than those of hydrodynamic ones, their angular distributions can be modeled by the same functions, so that they differ only quantitatively by the degree of mixing.
	The best-fit parameters to these models are listed in Table \ref{tab_summary}\footnote{We do not provide a best-fit value for $ f_c $ since the slower parts of the cocoon are yet to break out when the jet head reaches 2 breakout radii, and thus the values of $ f_c $ are expected to change considerably afterwards.}$^,$
	\footnote{We cannot provide best-fit values for the sGRB model $ \h $ since the numerical noise, which appears after the jet breakout from the ejecta, has a non-negligible contribution to the distributions.
		However, Figure \ref{fig:hg_axis}d shows that this jet features an utterly stable jet, more than hydrodynamic sGRBs and magnetized lGRBs, with no indication of a temporal evolution along the axis. That implies that magnetized sGRB jets dominate the outflow energy, with the weakest JCI and steepest power-law segments.}.
	
	The isotropic equivalent energy distribution of the lGRB magnetized jet of model $\Ce$, when the head reaches $2R_*$ is similar to that of the hydrodynamic sGRB jet, showing a relatively unmixed and energetic core with $\lambda \approx 0.3$ and a steep JCI power-law $ \delta \approx 3$. This is in contrast to the highly mixed hydrodynamic lGRB jet, $\A$, which shows almost no distinct jet core when the jet head is at $R_*=2$. The $u_\infty$ distribution of the magnetized lGRB jet, on the other hand, is more similar to that of the hydrodynamic lGRB jet, with $p_u \approx 2$. A comparison of the distributions of the hydrodynamic jets at $2R_*$ and $10R_*$ shows almost no evolution of the sGRB jet compared to a  significant evolution of the lGRB jet. The reason is that the mixing of the sGRB jet is low  at all times, with almost no evolution in time, while the mixing of the lGRB jet is very high upon breakout and it drops significantly at later times.
	In \S\ref{sec:distributions} we showed that magnetized lGRB jets show some reduction of mixing with time inside the star. This reduction is larger than the one seen in hydrodynamic sGRB jets but much lower than the one seen in hydrodynamic lGRB jets.
	It is therefore expected that at late times the JCI segment in model $ \Ce $ would become somewhat steeper with  lager values of $\delta$ and $p_u$. This result suggests that even a moderate level of subdominant magnetization can significantly alter the structure of lGRB jets, increasing significantly the energy in the jet core and reducing the JCI energy. This difference in the jet structure may be reflected in the observations, as we discuss next.

	\begin{table}
		\setlength{\tabcolsep}{14pt}
		\centering
		\begin{tabular}{ | l | c  c  c | }
			\hline
			Model & $ \lambda_2~(\lambda_{10}) $ & $ \delta_2~(\delta_{10}) $ & $ p_{u,2}~(p_{u,10}) $\\ \hline
			$ \A $ & 0.05 (0.13) & -- (1.8) & 1.7 (2.4) \\
			$ \G $ & 0.40 (0.38) & 3.0 (3.2) & 3.2 (3.2) \\
			$ \Ce $ & 0.32 & 2.8 & 2.2 \\\hline
		\end{tabular}
		
		\hfill\break
		
		\caption{
			A summary of the models characteristics: $ \lambda $ is the jet core energy to the total energy ratio, $ \delta $ is the power-law index in the angular energy distribution of the JCI. $ p_u $ is the power-law index in the angular distribution of the energy-weighted average of the four-velocity. Subscripts 2 and 10 reflect the location of the jet head, at 2 and 10 breakout radii, respectively.
			When the jet head reaches $ 2R_\star $ in model $ \A $ the cocoon dominates the JCI (see Figure \ref{fig:hydro_structure}a), and hence $ \delta_2 $ is not provided.
		}
		
		\label{tab_summary}
	\end{table}
	
	\subsection{Implications on the observed emission}
	\label{sec:emission}
	
	After breakout freshly injected jet material is free to accelerate once it exits the star and eventually radiates when its optical depth drops below $ \sim 1 $. The characteristics of the prompt and afterglow emissions will be governed by the mixing and the angular distribution of the radiating material, which is set during the passage through the confining medium. 
	The differences that we highlighted between hydrodynamic and weakly magnetized jets imply that even if the magnetic field is dynamically unimportant ($\sigma\ll1$), it may still influence the emission properties.
	
	The mixing can play a variety of roles when considering the prompt emission. \citet{Gottlieb2019b} showed that the photospheric emission from continuously injected hydrodynamic lGRB jets is highly efficient, unless the jet's average asymptotic Lorentz factor is $ \lesssim 100 $, much below current estimates. They found that the mixing in these jets leads to the formation of regions with different Lorentz factors inside the jet, which results in internal shocks and highly variable photospheric emission. These results are consistent with the prompt emission light curves.
	
	In our simulations of weakly magnetized jets, the flow is governed by hydrodynamic processes, so that the photospheric emission is expected to be efficient as well\footnote{The radiative efficiency from weakly magnetized jets may be even higher as less loading lowers the optical depth and places the photosphere at smaller radii.}.
	When considering the implications of boundary instabilities on the prompt emission light curve, two differences arise between magnetized and hydrodynamic lGRBs, which originate in the relation between mixing and temporal evolution.
	One on hand, GRB light curves show no indication for temporal evolution. However, all our lGRB simulations suggest that such evolution exists due to the change in mixing over time. Magnetized jets show less evolution in time and thus are more consistent with observations in this regard.
	On the other hand, GRB light curves are highly variable. The suppression of the baryon loading in magnetized jets leads to mild fluctuations in the Lorentz factor and thus only mild variability in the light curve, which is in some tension with observations and favors hydrodynamic models in this regard.
	Note that some magnetic jets, e.g. model $ \F $, exhibit both mixing and variation in time. Thus, similar characteristics of hydrodynamic jets are expected at first, before the mixing decreases and the jets become more stable
	(see e.g. bottom panel in Figure \ref{fig:hg_axis}).
	
	It appears that high variability, which is related to large mixing inevitably results in an undesirable evolution in time, since the mixing relaxes after the jet breakout.
	This problem may call for alternative variable engine models to account for the observed variability.
	\citet{Gottlieb2020a} recently found that modulations in engines of hydrodynamic lGRB jets result in intense mixing in the jet material launched during low-power episodes. Subsequently the heavy loaded jets cannot reach sufficiently high Lorentz factors and are incapable of generating the observed prompt emission. Magnetic fields may stabilize the jet enough to reduce the baryon load, while the temporal evolution will continue to be controlled by engine modulations. However, we caution that, since the mixing seen in the modulated hydrodynamic jet does not originate from the instabilities at the jet boundary, magnetic fields may not be able to considerably reduce it. Nevertheless, it is interesting to study whether such systems can produce variable and efficient $ \gamma $-ray emission.
	
	While the variation in the Lorentz factor affects mostly the prompt emission, the afterglow signal largely depends on the angular energy distribution of the jet-cocoon system (Figure \ref{fig:hydro_structure}).
	In GNB20 we pointed out that in typical hydrodynamic lGRBs most of the jet energy resides in the JCI ($\delta < 2 $) and hence both their on-axis and off-axis emissions have different forms from those of sGRBs (less mixed with typically $ \delta \approx 3 $).
	The JCI alters the jet emission in several ways. For an observer within the jet cone, the deceleration of the outflow to  Lorentz factors $<\theta_j^{-1}$ is accompanied by a steepening in the light curve power-law index known as the jet-break. For a lowly mixed jet the transition is sharp, while an energetic JCI will soften the transition making it very gradual.  
	An observer situated outside the jet cone of a lowly mixed jet, will see a prominent peak when the jet decelerates and the observer enters the emission cone of the jet core, which follows the analytic curve of a 'top-hat' jet \citep{Gottlieb2019a}. However, if $ \delta < 2 $, the JCI dominates at the time of the peak. After the peak a shallow drop is apparent, owing to the contribution of the jet core, and only at later times the power-law converges to the top-hat scenario. This implies that the analytic relations of off-axis top-hat jets do not apply for hydrodynamic lGRBs.
	We find that weakly magnetized jets (both lGRBs and sGRBs) exhibit similar behavior as hydrodynamic sGRBs and thus should obey the same 'top-hat' relations for the angular energy distribution after breakout.
	Therefore, weakly magnetized lGRB jets feature different emission afterglow light curves, mostly when seen off-axis, than hydrodynamic jets.
	This result implies that afterglow observations may provide us an opportunity to learn about the magnetic nature at the base of long GRBs.
	
	\section{Conclusions}
	\label{sec:conclusions}
	
	We study the effects of weak magnetic fields on the evolution of relativistic GRB jets. 
	Previous 3D numerical studies of long GRB jets
	\citep{Matsumoto2013a,Matsumoto2019,Gottlieb2019b} have shown that  hydrodynamic jets are prone to Rayleigh-Taylor and Richtmeyer-Meshkov instabilities that grow on the Jet-cocoon boundary surface above the collimation point. These instabilities erode the jet spine over time, leading to a diffused structure that separates the jet from the cocoon termed the {\it jet-cocoon interface} (JCI). In a companion paper, GNB20, we study the conditions leading to the formation of the JCI and the effect it has on the prompt and afterglow emission in hydrodynamic jets. We find that the structure of all hydrodynamic jets is composed of three regions with the following $\Eiso$ and $u_\infty$ angular distributions: (i) a jet core with flat distributions, (ii) JCI with power-law distributions, and the (iii) cocoon with an exponential cutoff $\Eiso$ distribution. The main factor that determines the  outflow's overall structure is the level of mixing that it suffers along its boundary, which  is expressed mostly in the values of the JCI power-law indices.
	
	In this work we show that subdominant ($ \sigma \ll 1 $) toroidal magnetic field can increase the jet stability against boundary layer instabilities	while avoiding global magnetic instabilities that grow in strong fields. We find that weakly magnetized jets also follow the same general structure that we find for hydrodynamic jets, where the reduced mixing is reflected in a less energetic JCI.
	
	In hydrodynamic lGRB jets, GNB20 find considerable instabilities that reach the jet axis and modify its composition. As a result, the final Lorentz factor of the jet core is reduced and most of the jet energy is transferred to the JCI. The result is a core with flat profiles of energy and four-velocity angular distributions, and a JCI with a rather shallow power-law distribution $\Eiso \propto \theta^{-\delta}$ with $\delta \approx 1-2$. The stochastic nature of the instabilities induces highly fluctuated jet baryon loading, which in turn translates to a large variability in the efficiency of the photospheric emission. The fluctuations in the baryon loading also facilitate internal shocks that lead to further dissipation and can significantly alter the prompt emission spectrum. In this paper we show that a toroidal magnetic field with $\sigma \gtrsim 10^{-2}$ is sufficient to considerably stabilize the boundary layer of typical lGRB jets. As a result, the jet core remains largely intact and a much smaller fraction of the jet energy is transferred to the JCI. This fraction decreases for increasing magnetic field strength. We were able to follow only the simulation with $\sigma = 10^{-2}$ for long enough times to obtain a reasonable approximation for the asymptotic structure of the outflow. We find that the JCI can still be approximated by power-law angular distributions of energy and asymptotic 4-velocity, but the inhibition of energy transfer from the jet core increases the energy distribution power-law index to $\delta \approx 3$. We also find that the stabilization of the jet significantly reduces the variability of the outflow. 
	
	Hydrodynamic sGRB jets are much more stable, due to a much lower density of the merger ejecta into which they are launched. Thus, while magnetized sGRB jets are more stable than hydrodynamic ones, the difference is not  significant. We were unable to obtain an approximation to the angular structure of weakly magnetized sGRB jet, but our simulations suggest that the value of $\delta$ in these jets is similar or even steeper than the one seen in hydrodynamic jets (where $\delta \approx 3$).

	The mixing and subsequent baryon contamination play a decisive role in shaping the light curve of the prompt GRB emission (\citealt{Gottlieb2019b}; GNB20). If the jet is continuously injected, an efficient photospheric emission emerges. High mixing introduces highly variable baryon loading which is translated to a variable photospheric emission and leads to internal shocks between sections with different loading in the jet.
	The reduced mixing in weakly magnetized jets increases the efficiency of the photospheric emission, but at the same time it yields a relatively smooth light curve and much weaker internal shocks. Intermittent hydrodynamic jets are subject to intense mixing and fail to generate prompt emissions \citep{Gottlieb2020a}. The effect of magnetic fields on intermittent jets is unclear since boundary instabilities is not the major source of the strong mixing seen in intermittent jets. Nevertheless, weakly magnetized intermittent jets may give rise to a variable prompt emission light curve. We address this issue in a future paper.
	
	The jet angular structure also affects the afterglow emission. Thus, the differences between hydrodynamic and magnetized lGRB jets should be manifested in the afterglow signature. When the JCI dominates the outflow energy, as in hydrodynamic lGRB jets, its contribution alters both the on-axis and off-axis emissions.
	The stabilization of the jet boundary by magnetic fields reduces the energy content in the JCI, making the afterglow emission of magnetized lGRBs behave as a classical top-hat jet when observed on-axis and as sGRBs (hydrodynamic and magnetized) when observed off-axis \citep{Gottlieb2019a}.
	
	\section*{Acknowledgements}
	
	The authors would like to thank A. Levinson and A. Phillipov for helpful comments and discussions.
	This research is partially supported by an ERC grant (JetNS) and an ISF grant (OG and EN). OB. and CS. were funded by an ISF grant 1657/18 and by an ISF (I-CORE) grant 1829/12.
	
	\section*{Data Availability}
	
	The data underlying this article will be shared on reasonable request to the corresponding author.
	
	\bibliographystyle{mnras}
	\bibliography{Magnetic_structure}
	
	\appendix
	
	\section{Convergence tests}
	\label{sec:convergence}
	
\begin{figure}
		\centering
		\includegraphics[scale=0.23]{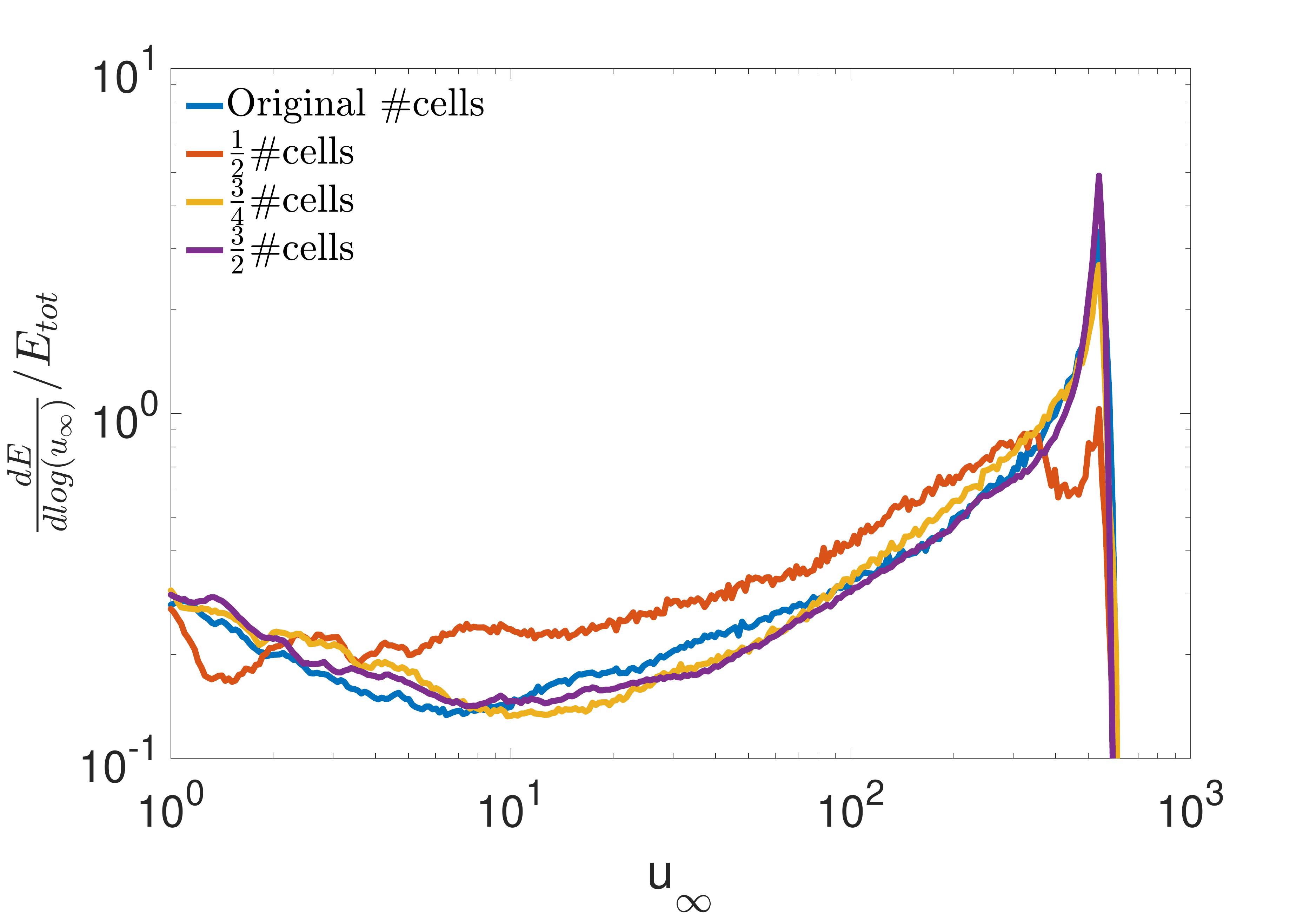}
		
		\caption[Convergence]{
			The logarithmic energy distribution as a function of $ u_\infty $ for model $ \C $ inside the star at different grid resolutions, normalized by the total energy of the original simulation.
			All runs have the same grid distribution up to a constant factor of 1.5.
		}
		\label{fig:convergence}
	\end{figure}

	We show that our results are independent of the grid setup with convergence tests of the degree of mixing in different resolutions. We perform simulations of the same setup of simulation $ \C $ with $ \frac{1}{2},\frac{3}{4} $ and $ \frac{3}{2} $ the number of cells in the original setup. We compare the energy distribution per a logarithmic scale of the four velocity of all runs in Figure \ref{fig:convergence}. One can see that while the simulation with half the number of cells is inconsistent with the original distribution, all the other runs agree with each other up to a factor of $ \sim 20\% $. We also verify that the pocket which has an important effect on the jet evolution has the same characteristics in the higher resolution runs.
	
	\section{The outer cocoon}
	\label{sec:outer_cocoon}
	
\begin{figure*}
		\centering
		\includegraphics[scale=0.39]{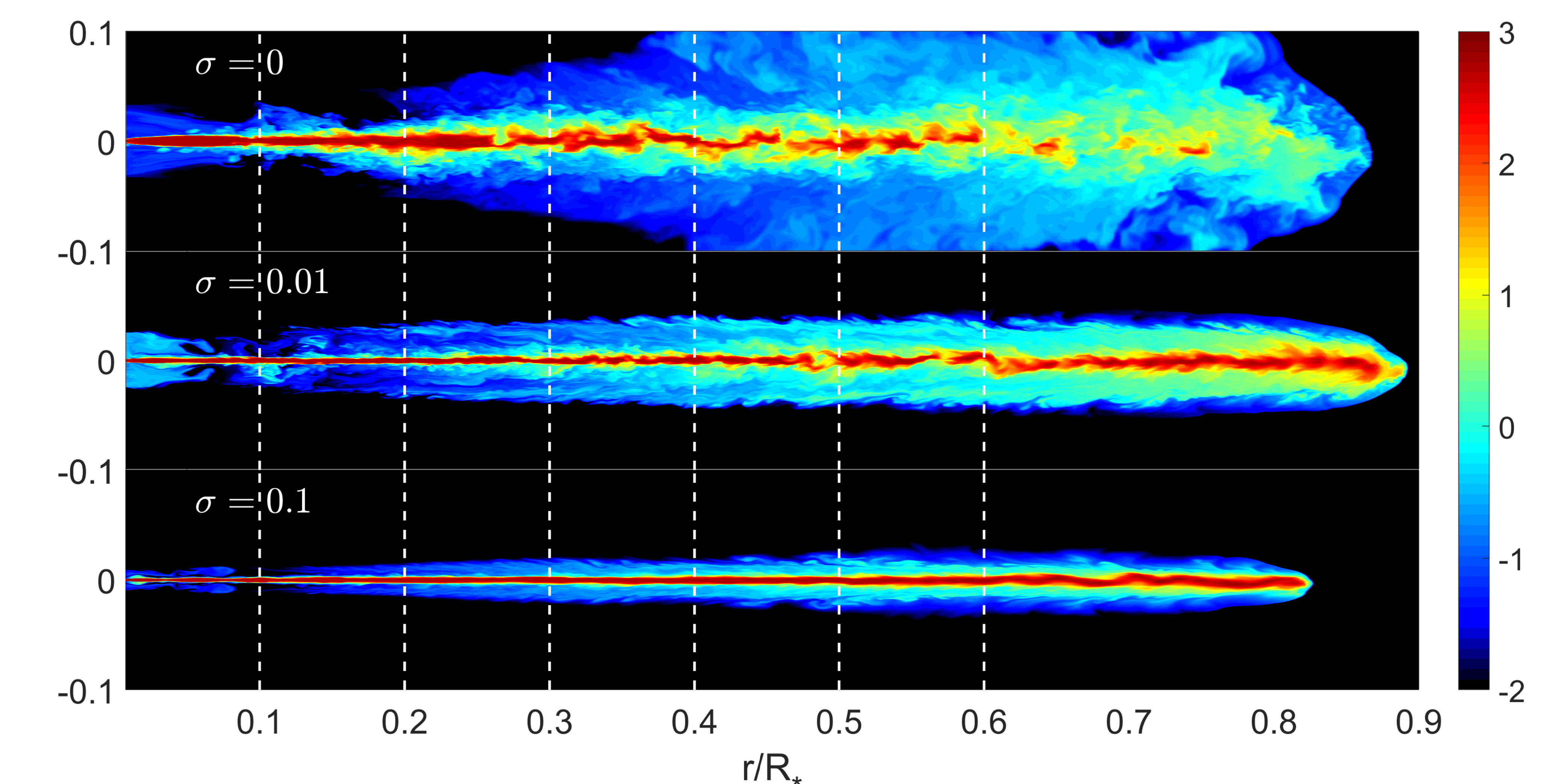}
		\includegraphics[scale=0.39]{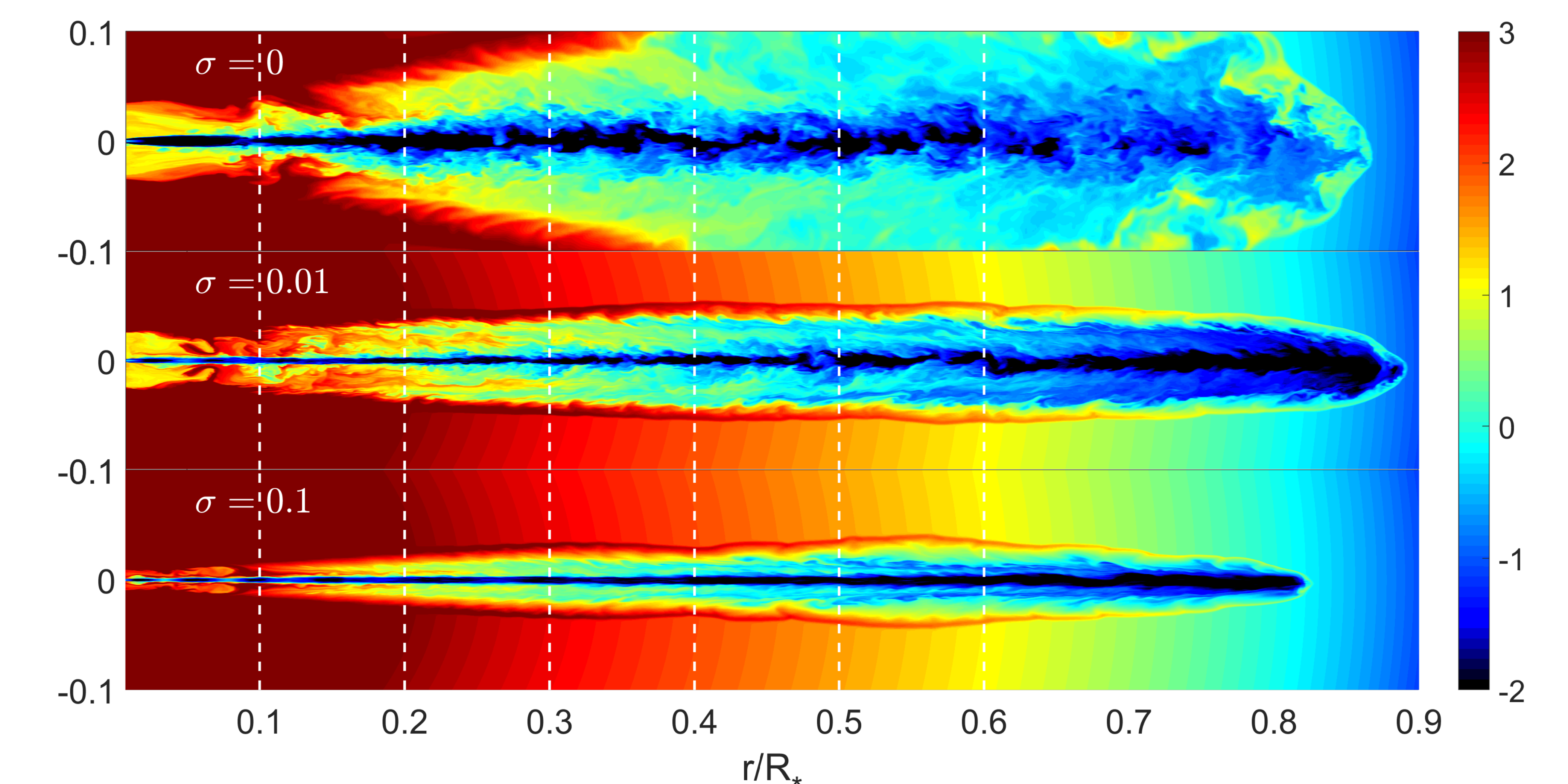}
		\includegraphics[scale=0.39]{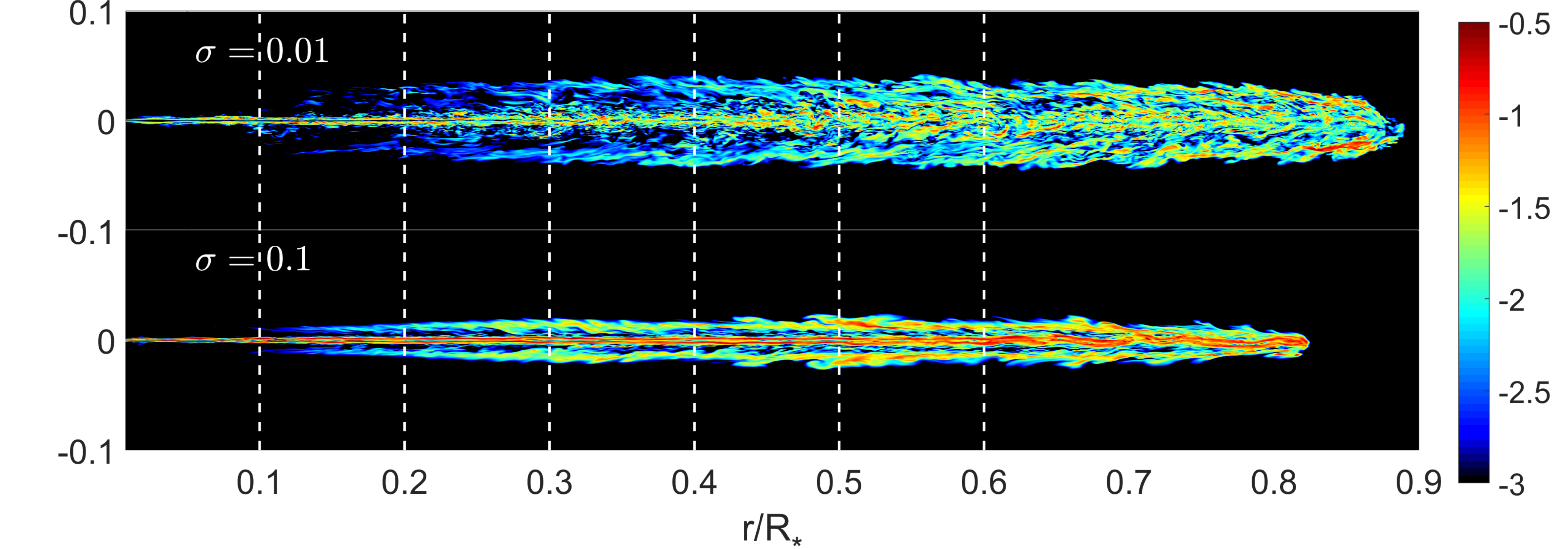}
		\caption[Extended figures]{
			A comparison of models $ \A, \C $ and $ \D $, similar to Figure \ref{fig:magnetic_cuts}, but with a logarithmic $ u_\infty $ (top) that is extended up to $ u_\infty = 10^{-2} $, so that the outer cocoon, characterized by $ u_\infty \lesssim 0.1 $ is also noticeable around the inner cocoon. Also shown are the logarithmic mass density for the three models (middle) and $ \sigma $ for the magnetized jets (bottom).
		}
		\label{fig:magnetic_cuts_extended}
	\end{figure*}

	We present more detailed maps that include the shocked medium in the outer cocoon. In the top panel of Figure \ref{fig:magnetic_cuts_extended} the outer cocoon can be identified as the region with $ u_\infty \lesssim 0.2 $. One can see that in the hydrodynamic jet the outer cocoon is much more extended than in the magnetic configurations. This is a direct result of the faster propagation of magnetized jets.
	In the middle panel we present the mass density to demonstrate once again that the heavy outer cocoon only plays a minor role in the magnetic cocoons. Finally, maps of $ \sigma $ for the magnetized jets are depicted in the bottom panel. It is shown that the value of $ \sigma $ is maximal and rather fixed along the jets, while the shocked jet material in the inner cocoons is also weakly magnetized.
	
	\label{lastpage}
\end{document}